\documentclass[%
 reprint,
superscriptaddress,
 amsmath,amssymb,
 aps,
]{revtex4-2}

\usepackage{inputenc}
\usepackage{graphicx}%
\usepackage{dcolumn}%
\usepackage{bm}%
\usepackage{comment}
\usepackage{physics}
\usepackage{amssymb}
\usepackage{amsmath}
\usepackage[hidelinks]{hyperref}%
\usepackage{dsfont}
\usepackage{xcolor}
\usepackage[caption=false, position=top]{subfig}
\usepackage{enumitem}
\usepackage{siunitx}
\usepackage{algorithm}
\usepackage{algpseudocode}

\newcommand{\boldbeta}{\ensuremath{\boldsymbol\beta}}
\newcommand{\boldgamma}{\ensuremath{\boldsymbol\gamma}}
\newcommand{\h}[1]{\ensuremath{\hat{#1}}}

\DeclareMathOperator{\sign}{sign}

\newcommand{\hhat}{\h{H}}

\newcommand{\hcost}{\ensuremath{\h{H}_{\mathrm{c}}}}
\newcommand{\hmix}{\ensuremath{\h{H}_{\mathrm{mix}}}}
\newcommand{\reducebar}{\ensuremath{\overline{\texttt{Reduce}}}}

\begin{document}

\preprint{APS/123-QED}

\title{Quantum-Informed Recursive Optimization Algorithms}%

\author{Jernej Rudi Fin\v{z}gar}%

\thanks{These authors contributed equally.}
\affiliation{BMW AG, Munich, Germany}
\affiliation{Technical University Munich, School of CIT, Department of Computer Science, Garching, Germany}

\author{Aron Kerschbaumer}

\thanks{These authors contributed equally.}
\affiliation{BMW AG, Munich, Germany}
\affiliation{Technical University Munich, School of CIT, Department of Computer Science, Garching, Germany}

\author{Martin~J.~A.~Schuetz}
\affiliation{Amazon Quantum Solutions Lab, Seattle, WA, USA}
\affiliation{AWS Center for Quantum Computing, Pasadena, CA, USA}

\author{Christian B. Mendl}
\affiliation{Technical University Munich, School of CIT, Department of Computer Science, Garching, Germany}
\affiliation{Technical University of Munich, Institute for Advanced Study, Garching, Germany}

\author{Helmut G. Katzgraber}
\affiliation{Amazon Quantum Solutions Lab, Seattle, WA, USA}

\date{\today}%

\begin{abstract}
We propose and implement a family of quantum-informed recursive optimization (QIRO) algorithms for combinatorial optimization problems.
Our approach leverages quantum resources to obtain information that is used in problem-specific classical reduction steps that recursively simplify the problem.
These reduction steps address the limitations of the quantum component (e.g., locality) and ensure solution feasibility in constrained optimization problems.
Additionally, we use backtracking techniques to further improve the performance of the algorithm without increasing the requirements on the quantum hardware.
We showcase the capabilities of our approach by informing QIRO with correlations from classical simulations of shallow circuits of the quantum approximate optimization algorithm (QAOA), solving instances of maximum independent set and maximum satisfiability problems with hundreds of variables.
We also demonstrate how QIRO can be deployed on a neutral atom quantum processor to find large independent sets of graphs.
In summary, our scheme achieves results comparable to classical heuristics even with relatively weak quantum resources.
Furthermore, enhancing the quality of these quantum resources improves the performance of the algorithms.
Notably, the modular nature of QIRO offers various avenues for modifications, positioning our work as a template for a broader class of hybrid quantum-classical algorithms for combinatorial optimization.
\end{abstract}

\maketitle

\section{Introduction}
\label{sec:introduction}

Quantum optimization has been identified as a promising area of research towards practical quantum advantage. On noisy intermediate-scale quantum (NISQ) devices, much effort has been dedicated to studying hybrid quantum-classical algorithms such as the quantum approximate optimization algorithm (QAOA)~\cite{Farhi2014-qaoa}. Importantly, QAOA is a \emph{local} algorithm. This means that at any constant circuit depth, only qubits that are separated by less than a certain distance in the interaction graph of the optimization problem are able to communicate. Together with the solution space properties of certain optimization problems such as the $\mathds{Z}_2$ symmetry, or the overlap gap property~\cite{gamarnik_overlap_2021}, locality was shown to severely limit the performance of QAOA~\cite{bravyi_obstacles_2020,farhi_quantum_2020-1, farhi_quantum_2020-2, Chou2021}.

Because introducing \emph{non-local} updates in quantum algorithms comes with additional hardware requirements, applying non-local updates classically has been proposed as an alternative in recursive QAOA (RQAOA)~\cite{bravyi_obstacles_2020, bravyi_hybrid_2022}. Here, values of variables are iteratively frozen by rounding the correlations between variables as measured in the quantum state prepared by QAOA. Hence, as variables are removed from the optimization problem the distances between nodes in the interaction graph are reduced iteratively. As such, the non-local effects introduced via the new connections between previously unconnected nodes counterbalance the locality inherent to QAOA.

\begin{figure}[b]
    \centering
    \includegraphics{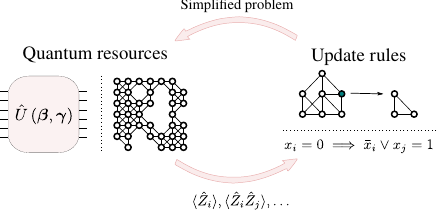}
    \caption{Schematic visualizing the core principles of the \emph{quantum-informed recursive optimization (QIRO)} algorithm presented in this work. Quantum resources (e.g., QAOA) are used to obtain information, e.g., in the form of one-point and two-point correlations, respectively.
    This information is then used to simplify the problem through problem-specific classical update rules.
    Here we exemplify the update rules with simple examples for the maximum independent set and satisfiability problems -- details on the update rules can be found in Section~\ref{sec:update-rules}. 
    The (simplified) problem obtained by means of the classical update rules is then used to restart the cycle. 
    The algorithm terminates when the problem has been fully simplified.}
    \label{fig:qiro-schematic}
\end{figure}

Building upon RQAOA, here we propose a family of hybrid quantum-classical algorithms, dubbed \emph{quantum-informed recursive optimization (QIRO)}. In QIRO, information generated by quantum resources is used to recursively reduce the size of the optimization problem by means of problem-specific classical optimization routines; see Fig.~\ref{fig:qiro-schematic} for a schematic illustration. This scheme allows us to leverage decades of research in (classical) combinatorial optimization~\cite{mcclean_low-depth_2021}, and tailor the classical subroutines to the particular optimization problem of interest, thereby enhancing the algorithm's performance.

Moreover, the inclusion of problem-specific update rules comes with additional benefits. It allows us to broaden the scope of our algorithm beyond gate-based architectures to analog devices, as shown in our experiments with the QuEra Aquila neutral atom analog quantum processor. This approach is reminiscent of previous approaches leveraging spin-freezing schemes on quantum annealers and in various Monte Carlo methods~\cite{karimi_effective_2017}. Furthermore, for problems with hard constraints, the update rules can offer options to enforce feasibility by design, even in the presence of noise.

Finally, we propose \emph{backtracking} as a strategy to further improve the performance of QIRO, by attempting to identify and correct non-ideal decisions made at earlier stages of the algorithm. Backtracking provides a way to enhance algorithmic performance without necessitating an increase in circuit depth, as is commonplace in quantum optimization algorithms.
Taken together, the robustness of QIRO, its applicability to analog devices, and the possibility to improve its efficacy without demanding better quantum resources, make QIRO a promising NISQ algorithm. 

In this work we provide concrete implementations for two paradigmatic NP-hard combinatorial optimization problems, namely, maximum independent set and maximum satisfiability. However, our work should be seen mainly as a proof-of-concept demonstration of a general template for designing hybrid quantum-classical algorithms for combinatorial optimization. We study the performance of QIRO by means of large-scale simulations of low-depth QAOA circuits with up to two hundred variables. Furthermore, we use QIRO to solve the maximum independent set problem on the QuEra Aquila neutral atom quantum device accessed via Amazon Braket. Finally, we compare the performance of QIRO against commonly used classical and quantum optimization techniques. This allows us to assess the role of the quantum resources used in QIRO, and showcase the value of classical subroutines.

The remainder of the paper is organized as follows. In Section~\ref{sec:problem-classes} we introduce the two problem classes studied in this work, in Section~\ref{sec:algorithm} we provide details about QIRO and in Section~\ref{sec:results} we study the performance of QIRO on the maximum independent set and maximum satisfiability problems. We discuss the implications of our findings in Section~\ref{sec:discussion} and we conclude by suggesting potential extensions to our framework in Section~\ref{sec:outlook}.

\section{Problem classes}
\label{sec:problem-classes}

Because \emph{problem-specific} update rules are at the core of QIRO, we first introduce the specific optimization problems considered in this work. We formalize these problems, and provide mappings to quantum Hamiltonians $\hcost$, which can in turn be implemented on different quantum devices~\cite{lucas_ising_2014, Glover2018}.

\subsection{Maximum Independent Set}
\label{sec:mis-intro}

The maximum independent set (MIS) problem is a paradigmatic NP-hard combinatorial optimization problem. It is of commercial relevance in areas such as network design or traffic optimization~\cite{mis_app}. For a graph $G=(V,E)$ with vertex set $V$ and an edge set $E$, the MIS problem amounts to finding the largest subset $S\subseteq V$ such that no two vertices in $S$ are adjacent [see Fig.~\ref{fig:example-graph} below for an example]. For a binary vector $\mathbf{x}\in\{0, 1\}^{|V|}$ finding the MIS is equivalent to finding the ground state of the following (classical) Hamiltonian:
\begin{equation}
\label{eq:mis-hamiltonian}
    H(\mathbf{x})
    =
    -\sum_{i\in V}x_i
    + 
    \lambda \!\!\!\sum_{(i, j) \in E} x_i x_j.
\end{equation}
Here, $\lambda > 1$ is a penalty term to enforce the independence constraint~\cite{lucas_ising_2014}. However, one should be mindful of the pitfalls of enforcing the independence constraint as a soft penalty term, because low-energy states (other than the ground states) of $H(\mathbf{x})$ might not necessarily obey the independence constraint. Performing the mapping ${z_i = 2 x_i - 1}$ to Ising variables $z_i = \pm 1$, and promoting the spin variables to quantum operators ${z_i \rightarrow \h{Z}_i}$ translates Eq.~\eqref{eq:mis-hamiltonian} to a quantum Hamiltonian suitable for quantum devices~\cite{lucas_ising_2014}.

Moreover, the MIS problem on a family of graphs called unit disk graphs (UDGs) has been shown to have a natural mapping to neutral atom quantum devices~\cite{pichler_quantum_2018, ebadi_quantum_2022}. There, the Rydberg blockade mechanism---which prevents simultaneous excitations of nearby atoms---enforces the independence constraint. Finding the MIS of a general UDG, despite being a specialized family of graphs, remains an NP-Hard problem~\cite{pichler_quantum_2018} and has commercial applications (e.g., in network design~\cite{wurtz_industry_2022_1}). For additional details we refer the reader to Appendix~\ref{app:quera-details}.

\subsection{Maximum Satisfiability}

The satisfiability problem (SAT) is arguably one of the best-studied combinatorial optimization problems and was the first problem proven to be NP-complete~\cite{cook_1971}. Given a propositional logic formula $\phi$ of $n$ Boolean variables $x_i\in\left\{0,1\right\}$, SAT is a decision problem which asks for the existence of an assignment of the variables $x_i$ that satisfies $\phi$. Each clause can then be written as 
$$\Tilde{C}_j=\bigvee_{k=1}^{K_j} \ell_{j_k},$$
where literals $\ell_{j_k}$ are either a variable $x_{j_k}$ or its negation $\Bar{x}_{j_k}$, for $j_k \in \left\{1,\dots,n\right\}$, and $K_j$ is the length of the $j$-th clause. For a fixed $K_j=K$ the corresponding problem of whether or not the Boolean formula with $m$ clauses
$$
\phi
=
\bigwedge_{j=1}^{m}
\Tilde{C}_j
$$
has a satisfying assignment, is referred to as $K$-SAT. 

Here, we focus on the optimization version of satisfiability, referred to as maximum satisfiability (MAX-SAT). In MAX-SAT, the goal is to find an assignment of the variables $x_i$ that minimizes the number of violated clauses. The MAX-$K$-SAT problem deals with propositional logic formulae with \emph{at most} $K$ literals per clause. 
An important parameter of a (MAX-)$K$-SAT problem is the clause-to-variable ratio $\alpha:=m/n$, which strongly influences the algorithmic hardness of random problem instances~\cite{sat-phase-trans}. Research indicates that there is a phase transition in computational complexity~\cite{zeng_schedule_2016}. While for $K$-SAT an easy-hard-easy transition is observed~\cite{Mezard2002, Sleegers2020}, MAX-$K$-SAT exhibits a transition from easy underconstrained to hard overconstrained problems at a certain $\alpha_c$~\cite{Ochoa2020}. In what follows we limit ourselves to $K=2$ (clauses contain either one or two literals), as MAX-$2$-SAT is already NP-hard~\cite{garey_johsnon_np}, despite its decision version ($2$-SAT) being solvable in linear time~\cite{2sat-proof}. Here, the phase transition occurs at $\alpha_\text{c} = 1$~\cite{Chvatal1992}. 

We use the fact that for Boolean (binary) variables $\Bar{x}_i\equiv (1-x_i)$, solving MAX-$2$-SAT for a given formula $\phi$ is equivalent to finding a ground state of the following (classical) Hamiltonian:
\begin{equation}
\label{eq:sat-hamiltonian}
    H(\phi)
    =
    \sum_{j=1}^{m}
    C_j
    =
    \sum_{j=1}^{m}
    \prod_{k=1}^{K_j}
    (1-\ell_{j_k}).
\end{equation}
It is important to keep in mind that $\ell_{j_k}$ are nothing but the Boolean variables $x_{j_k}$ (or their negations $\Bar{x}_{j_k}$). Hence, by using the transformation into Ising variables outlined in the case of the MIS problem in Section~\ref{sec:mis-intro}, we can likewise encode the MAX-2-SAT problem onto quantum devices.

\section{Algorithm}
\label{sec:algorithm}

We now provide details about QIRO. Essentially, in QIRO quantum resources are utilized to inform update rules that recursively simplify the problem at hand (see Fig.~\ref{fig:qiro-schematic}). The idea is a generalization of previous work, in particular RQAOA~\cite{bravyi_obstacles_2020, bravyi_hybrid_2022} (also described in Appendix~\ref{app:rqaoa}). However, in contrast to RQAOA, the update steps used in QIRO are problem-specific. This comes with several advantages and facilitates cross-pollination with ideas from classical optimization~\cite{mcclean_low-depth_2021}.

At each iteration of the QIRO algorithm, for a given problem Hamiltonian we first prepare a low-energy quantum state, generically in the form of a superposition of low-energy candidate solutions to the optimization problem. Next, we use this quantum state to extract information (e.g., correlations between variables), which is then passed to a classical update step with the goal of simplifying the problem. We design this update step to complement the quantum part of the algorithm by performing operations better suited to classical hardware. This may include non-local steps that address the identified limitations of (local) quantum algorithms, or enforcing hard constraints. As such, the classical update step (and the specific design thereof) is crucial to the performance of QIRO, and is typically problem-specific. As such, it might need to be modified by the end user, depending on their needs. However, templates for the update steps may be borrowed from the literature on classical combinatorial optimization~\cite{Papadimitriou1998-pc}. The procedure of quantum state preparation and the subsequent classical update step is repeated until the size of the problem is reduced sufficiently, such that, for example, we are able to solve it exactly by a brute-force search or other means if the reduction does not yield problem sizes amenable to exact solvers. Finally, we propose the use of backtracking to identify and rectify problem reductions that led to an unfavorable outcome, thus providing an additional strategy to obtain enhanced solutions without requiring deeper quantum circuits.

In what follows, we provide details on the individual components of QIRO. We begin by describing the quantum state preparation methods utilized in this work, followed by the proposed update rules for the MIS and MAX-2-SAT problems. Next, we introduce the general strategy of backtracking and describe the particular implementation used in this work. Finally, we combine the building blocks to present the complete QIRO algorithm.

\subsection{Quantum State Preparation}
\label{sec:quantum-state-preparation}

It is desirable that the quantum state which is prepared at the beginning of each iteration has a low expectation value with respect to the cost Hamiltonian $\hcost$. Therefore, we hope that the information guiding the classical reduction steps is extracted from (a superposition of) configurations corresponding to good candidate solutions to the optimization problem. A plethora of protocols for preparing low-energy states may be found in the literature; in this manuscript we consider QAOA~\cite{Farhi2014-qaoa} and adiabatic state preparation~\cite{farhi_aqc, Kadowaki_1998}, as outlined next.

\subsubsection{QAOA}
\label{qaoa-explanation}

Let us first introduce QAOA. For both the MAX-2-SAT and MIS problems considered in this manuscript, the cost Hamiltonian can be written in the general (Ising) form $\hcost=\sum_{i,j}J_{ij}\h{Z}_i\h{Z}_j + \sum_i h_i\h{Z}_i$, where we have omitted any constant terms. Using the cost Hamiltonian $\hcost$ from Section~\ref{sec:problem-classes}, and defining the mixer Hamiltonian as $\hmix = -\sum_{i=1}^{n} \h{X}_i$, the QAOA circuit is given by an alternating application of the time evolution generated by these two operators. Specifically, taking the initial state as $\ket{\psi_0}:=\ket{+}^{\otimes n}$, the state produced with QAOA at depth $p$ is $$\ket{\psi(\boldbeta, \boldgamma)} =e^{- i \beta_p \hmix} e^{- i \gamma_p \hcost} \cdots e^{- i \beta_1 \hmix} e^{- i \gamma_1 \hcost} \ket{\psi_0}.$$ The $2p$ parameters $(\boldbeta, \boldgamma)$ are then determined by a classical optimization routine such that $\bra{\psi(\boldbeta, \boldgamma)}\hcost\ket{\psi(\boldbeta, \boldgamma)}$ is minimized.

In this work, we predominantly focus on the lowest QAOA depth ($p=1$), with the goal to establish a lower bound for the performance of QIRO. It is natural to expect that the improved quality of the quantum information extracted from higher depth $p>1$ QAOA circuits (in the absence of noise) would enhance the performance of the algorithm. 
Our $p>1$ experiments on small problem instances shown in Appendix~\ref{app:higher-depth-qaoa} corroborate this intuition. 

For the state produced by QAOA at $p = 1$ the expectation values of the single- and two-body terms comprising $\hcost$ can be calculated classically in polynomial time, enabling us to perform numerical simulations of systems with several hundred variables~\cite{bravyi_obstacles_2020, bravyi_hybrid_2022,  Ozaeta2022-journal}. We note that there is no method known (to us) how this could be done efficiently for $p > 1$. For details about our implementation we refer the reader to Appendix~\ref{app:qaoa}.

\subsubsection{Adiabatic state preparation}

We next turn or attention to adiabatic state preparation. We present a general description of the protocol and defer the details pertaining to our implementation of the MIS problem on neutral atom quantum processors to Appendix~\ref{app:quera-details}.

At the beginning of an adiabatic protocol, we start in an easy-to-prepare ground state of a simple Hamiltonian, e.g., $\hmix$. We then slowly transform the Hamiltonian from $\hmix$ to $\hcost$. If the rate of change is slow enough compared to the smallest instantaneous spectral gap (between the instantaneous ground and first excited states), the adiabatic theorem guarantees that at the end of the protocol, the ground state of $\hcost$ is  obtained~\cite{Kadowaki_1998, farhi_aqc}. Inspired by adiabatic protocols, the quantum annealing computational paradigm has been subject of widespread interest in recent decades~\cite{hauke_perspectives_2020}. In quantum annealing one forgoes the strict requirement of adiabaticity and, in general, uses analog protocols to prepare near-ground states. These can, in turn, be used to inform the classical update steps.

\subsection{Update Rules}
\label{sec:update-rules}

We now discuss example update rules for the considered optimization problems. We begin by preparing a low-energy quantum state $\ket{\psi}$ from which information is distilled.
In QIRO this information comes in the form of one-point $\bra{\psi}\h{Z}_i\ket{\psi}$ and two-point correlations $\bra{\psi}\h{Z}_i\h{Z}_j\ket{\psi}$, which we store in the matrix $M$ (see Algorithm~\ref{alg:update-step}). Potential generalizations, such as using expectation values of other observables, are left for future research.

If the lowest-depth ($p=1$) of QAOA is used to generate these correlations, the only nonzero values of $M$ in an ideal numerical simulation as described in Appendix~\ref{app:qaoa} (i.e., free of shot and environmental noise) correspond to pairs of variables that are connected in the original interaction graph, i.e., to pairs of variables with nonzero coupling coefficients $J_{ij}$. Therefore, we design our update rules to only consider correlations between such connected variables. Inspired by RQAOA~\cite{bravyi_obstacles_2020, bravyi_hybrid_2022}, we next use these correlations to successively simplify the problem with each update step. However, in contrast to RQAOA, the QIRO update steps are problem-specific in that they exploit the structure of the given optimization problem. As such, at any stage of the algorithm, the problem that remains to be solved is a valid instance of the same class of optimization problems. This will be crucial when deploying QIRO to solve the MIS problem on neutral atom quantum hardware.

In the two specific cases detailed in the remainder of this section, only the largest correlation in terms of its absolute value is used to inform the update step, with ties broken at random. Intuitively, the largest correlation corresponds to a variable (pair) which appears in a certain configuration with high probability in the low-energy state $\ket{\psi}$. Thus, it is reasonable to use this information to inform the update step. For example, if a large positive correlation $\langle\hat{Z}_i\hat{Z}_j\rangle\approx 1$ is measured, it is likely that most of the low-energy configurations have $x_i$ and $x_j$ take the same value. Generalizations of this deterministic scheme to randomized (e.g., evolutionary) schemes will be studied in the future.

As we lay out next, these update steps are carried out in the original formulation of the respective optimization problems, facilitating the adoption of techniques from the extensive literature on classical combinatorial optimization. We call this update step \texttt{Reduce} and describe it in Algorithm~\ref{alg:update-step}.
\begin{algorithm}[H]
\caption{Reduce}
\label{alg:update-step}
\hspace*{\algorithmicindent} \textbf{Input:} Problem $P$, partial solution $S$. \\
\hspace*{\algorithmicindent} \textbf{Output} Simplified prob. $P^{\prime}$, updated partial solution $S^{\prime}$.
\begin{algorithmic}[1]
\State Prepare low-energy quantum state $\ket{\psi}$.
\State Store correlations in $M\in\mathbb{R}^{n\times n}$:\newline%
$\forall i \in [n]: M_{ii} = \bra{\psi} \h{Z_i}\ket{\psi}$,\newline
$\forall (i, j) \in [n] \times [n] \text{~s.t.~} J_{ij}\neq 0: M_{ij} = \bra{\psi} \h{Z}_i \h{Z}_j \ket{\psi}$. %
\State $P^{\prime}, S^{\prime} \gets \mathrm{Simplification}(P, S, M)$ \newline
\Comment{Problem-specific simplification is informed by the correlations stored in $M$.}
\State \Return $P^{\prime}$, $S^{\prime}$
\end{algorithmic}
\end{algorithm}
The problem-specific updates employed for the optimization problems considered in this work constitute the \texttt{Simplification} function in Algorithm~\ref{alg:update-step}, and are described next. However, we emphasize that the design choices we make are by no means the only option, with many modifications and extensions possible.

\subsubsection{MIS Simplification}
\label{sec:mis-update-rules}

\begin{figure}
\includegraphics{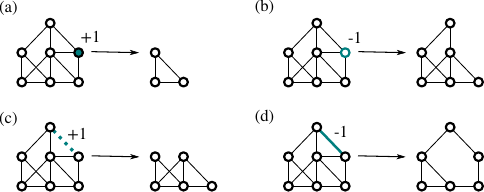}
\caption{MIS update rules described in the main text carried out on an example graph. Panels (a) and (b) show reduction rules for the case of a positive and negative \emph{one-point} correlation, respectively. Analogously, panels (c) and (d) show the respective reduction rules for the case of a positive and negative \emph{two-point} correlation.
}
\label{fig:mis-reductions}
\end{figure}

At the beginning of the \texttt{Simplification} routine for MIS, we find the entry of $M$ with the largest absolute value. Depending on the sign of the correlation with the largest absolute value and on whether it lies on the diagonal (corresponding to a one-point correlation) or not (corresponding to a two-point correlation), we perform a different simplification.
The simplifications are designed such that the correlation-informed reductions are consistent with the independence constraint.
The four cases are visualized in Fig.~\ref{fig:mis-reductions} and described next:
\begin{enumerate}[label=(\alph*)]
    \item If $M_{ii} \geq 0$ was selected, we set the $i$-th vertex to be in the independent set (IS). We then remove all vertices connected to the $i$-th node from the graph, as including them would violate the independence constraint.
    \item If $M_{ii} < 0$ was selected, we remove the $i$-th vertex from the graph.
    \item If $M_{ij} > 0$ [for $(i, j)\in E$] was selected, we remove both nodes from the graph -- this is because the only positively correlated assignment of \emph{connected} vertices consistent with the independence constraint is not to include either node in the IS.
    \item If $M_{ij} < 0$ [for $(i, j)\in E$] was selected, we remove nodes which are connected to both node $i$ and node $j$ simultaneously, i.e., we remove every node $k$ for which $(k,i)\in E$ and $(k, j)\in E$. Intuitively, we know that if the variables are negatively correlated, one of them will be assigned to be in the IS. Thus, we only remove variables which are connected to both of them simultaneously.
\end{enumerate}

If no nodes were removed from the graph using the described simplifications [only possible in the case (d)], we repeat the procedure using the next largest correlation in terms of its absolute value. Finally, we identify all connected components (i.e., connected subgraphs that are not part of any other connected subgraph) of the graph. If the connected component contains less than a certain number of vertices $n_{\mathrm{c}}$ (for concreteness, we use $n_{\mathrm{c}}=15$ here), we compute its maximum independent set by a brute-force search, and remove it from the graph. This step is mostly performed in an effort to save resources when the algorithm is deployed on real quantum hardware.

Crucially, the rules delineated above explicitly ensure that the independence constraint is obeyed. Hence, the solutions obtained by repeated applications of \texttt{Simplification} are definitely feasible. This holds true irrespective of the quality of the (quantum) information used to perform the simplification, thus bestowing the scheme with the resilience required in the NISQ era. However, although the update rules improve the performance of the algorithm by ensuring solution feasibility, the quality of the obtained solutions ultimately hinges on the quality of the supplied correlations.

Interestingly, in contrast to the RQAOA update rules the QIRO MIS update rules proposed here are \emph{local} according to the definition of locality in Section~\ref{sec:introduction}, as no new connections between variables are created. However, the RQAOA update rules lead to intermediate Hamiltonians that generically do not take the form of an MIS Hamiltonian as in Eq.~\eqref{eq:mis-hamiltonian}. Conversely, at each stage of the QIRO algorithm the reduced problem is a valid MIS problem (see also Fig.~\ref{fig:mis-reductions}). Thus, if one can encode the original MIS problem onto an analog quantum device (e.g., a neutral atom quantum processor), the same can be done with intermediate, simplified problems.
As such, one might be able to harness the nonlocal effects generated by the quantum many-body time evolution on a neutral atom quantum processor, circumventing the issues of local quantum algorithms. Previously, such adiabatic protocols have been shown to generate solutions surpassing those obtained by running QAOA, even at depths higher than $p=1$ used in this manuscript~\cite{ebadi_quantum_2022}.

\subsubsection{MAX-2-SAT Simplification}
\label{sec:m2s-update-rules}

In the case of MAX-2-SAT the \texttt{Simplification} step likewise begins by identifying the entry of $M$ with the largest absolute value. Depending on whether the largest value corresponds to a one-point or a two-point correlation, we apply a different reduction:
\begin{enumerate}[label=(\alph*)]
    \item If $M_{ii}$ was selected, we set the variable $x_i\rightarrow [\sign(M_{ii}) + 1]/2$, i.e., a positive (negative) one-point correlation corresponds to assigning $x_i$ to TRUE (FALSE). 
    \item If $M_{ij}$ (for $i<j$) was selected, we replace $x_i\rightarrow x_j$ if $\sign(M_{ij})=1$, and $x_i\rightarrow \Bar{x}_j$ if $\sign(M_{ij})=-1$.
    Intuitively, this update step captures the relationship between variables $x_i$ and $x_j$ as inferred from the correlation. If there is a positive (negative) correlation, we assign $x_i$ the same (opposite) value as $x_j$.
\end{enumerate}

We note that only the two-point correlation reduction step described in (b) constitutes a nonlocal update because it potentially introduces a new link between previously unconnected variables. Because we only consider formulae in conjunctive normal form (i.e., conjunctions of disjunctions), setting a certain variable to the value required by the clause is sufficient to satisfy the clause (see the example update step in Fig.~\ref{fig:qiro-schematic}). Hence, we can remove the clause from consideration, thus simplifying the problem. Specifically, all clauses in which a given literal evaluates to TRUE can be removed. Conversely, if the literal evaluates to FALSE, only the literal itself can be removed from the clause. If a certain clause becomes empty (i.e., if both literals were removed from it), it corresponds to an \emph{unsatisfied} clause.

Next, we apply inference rules, which is a standard technique in the field of satisfiability solvers~\cite{Abrame2015, Li2021a}. Inference rules deduce information from the structure of the current Boolean formula, simplifying it by assigning values to certain variables and thus speeding up further computation. The rules guarantee that the optimal solutions of the current and simplified problem are of equal quality. The chosen inference rules (e.g., the pure literal rule~\cite{Borchers1998}) are conceptually simple rules that incur a small computational overhead.
Further details on the specific inference rules we employ are provided in Appendix~\ref{app:m2s-update}. Finally, if the number of remaining variables falls below $n_{\mathrm{c}}$ (we use $n_{\mathrm{c}}=10$ for MAX-2-SAT), the solution of this simplified problem is found by a brute-force search of the remaining solution space. 

\subsection{Quantum Informed Recursive Optimization Algorithm}

We now possess all of the ingredients for the quantum-informed recursive optimization (QIRO) algorithm. At each step of QIRO, we begin by preparing a quantum state (as described in Section~\ref{sec:quantum-state-preparation}). We proceed by using the correlation information by following the problem-specific prescriptions from Section~\ref{sec:update-rules}. The simple end-to-end QIRO procedure is summarized in Algorithm~\ref{alg:qiro}. 
\begin{algorithm}[H]
\caption{Quantum-informed recursive optimization (QIRO)}\label{alg:qiro}
\hspace*{\algorithmicindent} \textbf{Input:} Problem $P$. \\
\hspace*{\algorithmicindent} \textbf{Output} Complete solution $S$.
\begin{algorithmic}[1]
\State Initialize empty solution $S$.
\While{$\mathrm{size}(P) > 0$}
\State  $P, S \gets \text{Reduce}(P, S)$\Comment{See Algorithm~\ref{alg:update-step}.}
\EndWhile
\State \Return $S$ \Comment{Return the solution.}
\end{algorithmic}
\end{algorithm}

\subsection{Backtracking}
\label{sec:backtracking}

We now turn our attention to backtracking. QIRO---as specified in Algorithm~\ref{alg:qiro}---is a polynomial-time algorithm. Therefore, under common complexity theoretical assumptions, there exist instances of NP-hard problems that QIRO will not be able to solve to optimality. This means that some of the reduction steps performed by QIRO to arrive at a candidate solution might be wrong. A similar observation was made for RQAOA in Ref.~\cite{patel_reinforcement_2022_1}, prompting the authors to consider reinforcement learning to steer RQAOA update steps. Here, we propose backtracking to derive improved solutions starting from an initial candidate solution obtained by QIRO. Through backtracking, we obtain a method to refine solutions by using the same quantum resources more frequently. This approach stands in stark contrast to most other quantum optimization algorithms, which usually require enhanced quantum resources (e.g., increasing the depth $p$ in QAOA) to deliver improved solutions. 

In this work, we only employ this strategy for the MAX-2-SAT problem.
However, extending this approach to other optimization problems such as the MIS problem is straightforward.

\begin{figure}
\includegraphics{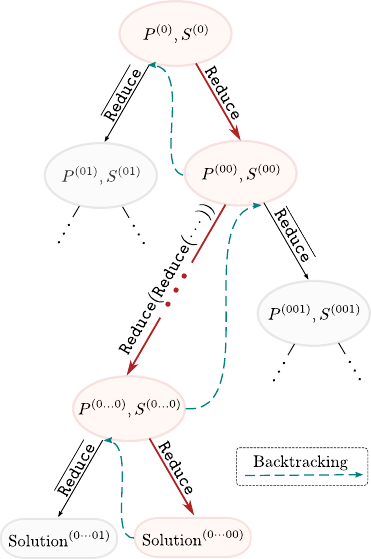}
\caption{
Schematic visualization of the backtracking procedure used to obtain improved solutions. Starting from the original problem $P^{(0)}$ and an empty solution $S^{(0)}$, we perform a series of reductions (red arrows \& nodes) that recursively simplify the problem, until the first candidate solution ($\mathrm{Solution}^{(0\cdots00)}$) is reached. This corresponds to applying QIRO as outlined in Algorithm~\ref{alg:qiro}. Backtracking, as visualized by the dashed teal arrows, then amounts to revisiting reduction steps made along the path. At the chosen ancestor node we have backtracked to, we can then make an alternative reduction, here denoted by \reducebar~(see black arrows in Fig.~\ref{fig:search-tree}). Concretely, \reducebar~amounts to reversing the initially made decision, not needing extra calls to the quantum device. In this work, we limit ourselves to making a binary decision at each node. As such, we can label the reduced problem and partial solution at each node with a binary string encoding the path connecting the node to the original problem node. Each $0$ ($1$) in the superscript labelling $P$ and $S$ corresponds to one application of $\texttt{Reduce}$ (\reducebar) starting from the original problem. 
}
\label{fig:search-tree}
\end{figure}

As visualized in Fig.~\ref{fig:search-tree} the simplification process of QIRO can be represented by means of a search tree. Decisions (i.e., update steps) made by the algorithm correspond to arrows in the search tree, which lead to simplified problems represented by nodes. Following the update rules in Section~\ref{sec:update-rules} (i.e., successively applying \texttt{Reduce}) traces out a path through the search tree (see red arrows and nodes in Fig.~\ref{fig:search-tree}), which eventually leads to a candidate solution (see $\mathrm{Solution}^{(0\cdots00)}$ in Fig.~\ref{fig:search-tree}). Here, the superscript $(0\cdots00)$ indicates that this is the solution obtained in the initial passage through the search tree.

We would like to consider alternative solutions, obtainable by deviating from the initial path through the search tree which led to the initial candidate solution $\mathrm{Solution}^{(0\cdots00)}$. To this end, we define the \reducebar~function, which reverses the decision previously made at the revisited node (i.e., partial problem and solution). We emphasize that this initial reversal of the previously made decision requires no additional measurements from the quantum device -- we simply perform the opposite reduction to the one we performed initially. For example, if at first the decision was made to assign the value of a variable $x_i=\mathrm{TRUE}$, we reverse that decision and set $x_i=\mathrm{FALSE}$ instead. Thus, we diverge from the original path through the search tree as visualized in Fig.~\ref{fig:search-tree}. We note that \reducebar~requires the information about the solution it is deviating from in order to be able to perform the reversed decision.

In principle, exploring the entire search tree essentially corresponds to a brute-force search of the complete solution space. Because we are interested in designing heuristic algorithms that run in polynomial time, we limit the permitted exploration. We find that a simple rule for which the number of QAOA circuit executions scales quadratically in the problem size is sufficient to produce satisfactory results for the considered problem sizes of MAX-2-SAT. However, more (or less) sophisticated approaches can be designed and tailored to the particular needs and resources of the end user. Next, we describe the specific backtracking strategy used in our simulations and refer the reader to Algorithm~\ref{alg:qiro+bt} for the pseudocode.

\begin{algorithm}[H]
\caption{QIRO + Backtracking (QIRO + BT)}\label{alg:qiro+bt}
\hspace*{\algorithmicindent} \textbf{Input:} Problem $P$.\\
\hspace*{\algorithmicindent} \textbf{Output} Complete solution $S$.
\begin{algorithmic}[1]
\State Initialize empty solution $S$.
\State $L\gets \{\}$ \Comment{List to store partial problems \& solutions.}
\While{$\mathrm{size}(P) > 0$}
\State Append $(P, S)$ to $L$.
\State  $P, S \gets \text{Reduce}(P, S)$\Comment{See Algorithm~\ref{alg:update-step}.}
\EndWhile
\For{$P^{\prime}, S^{\prime}$ in $L$}\Comment{Backtracking.}
\State $P^{\prime}, S^{\prime}\gets \overline{\text{Reduce}}\left(P^{\prime}, S^{\prime}; \mathrm{Solution}^{(0\cdots00)}\right)$\newline\Comment{Make alternative decision.}
\While{$\mathrm{size}(P^{\prime}) > 0$}
\State  $P^{\prime}, S^{\prime} \gets \text{Reduce}(P^{\prime}, S^{\prime})$
\EndWhile
\State Store the obtained solution.
\EndFor
\State \Return Best solution found.
\end{algorithmic}
\end{algorithm}

First, we generate the initial candidate solution $\mathrm{Solution}^{(0\cdots00)}$ by successively applying \texttt{Reduce} as visualized in Fig.~\ref{fig:search-tree}. Put differently, we first apply QIRO \emph{without} backtracking. Next, we backtrack to each of the ancestor nodes of the initial candidate solution (as visualized by dashed teal lines in Fig.~\ref{fig:search-tree}). Each of the revisited parent nodes corresponds to a different simplified problem and a corresponding partial solution. The decision made in the initial passage through the search tree is then reversed (by applying \reducebar), whereby a new node in the search tree (e.g., $P^{(01)}, S^{(01)}$ in Fig.~\ref{fig:search-tree}) is created. Here, each $0$ ($1$) in the superscript labelling $P$ and $S$ corresponds to one application of $\texttt{Reduce}$ (\reducebar) starting from the original problem. Subsequently, if the problem has not been fully reduced yet, from this new node a new candidate solution is generated by repeatedly applying \texttt{Reduce}. In contrast to the initial application of \reducebar~the subsequent applications of \texttt{Reduce} require additional calls to the quantum device -- see the definition of \texttt{Reduce} in Algorithm~\ref{alg:update-step}. In other words, each revisited node leads to a new candidate solution that is identical to the initial candidate solution in the values of the variables fixed before the revisited node. From the revisited node onward the new candidate solution differs from the initial solution. Finally, the output of the algorithm is the best candidate solution found in this way. We note that the order in which the parent nodes are revisited is not important. 

\section{Results}
\label{sec:results}

Next, we present the results obtained by applying QIRO to random instances of the MIS and MAX-2-SAT problems. In numerical simulations we compare the performance of quantum algorithms (QIRO, RQAOA) and two prominent classical heuristics (simulated annealing, greedy algorithms). Of course more complex tailored classical approaches can be used, such as the parallel tempering~\cite{Hukushima_1996,Earl2005,Roma2009}, borealis~\cite{zhu_borealisgeneralized_2020}, and extremal optimization~\cite{Boettcher_2003} heuristics, as well as a plethora of exact solvers benchmarked at the annual (MAX-)SAT competition~\cite{sat2023}.

Efficient classical simulations of $p = 1$ QAOA allow us to study the system size scaling on problem instances with various connectivities of the underlying interaction graph.
Results using the lowest depth of QAOA should mainly be interpreted as a lower bound on the performance of QIRO.
Furthermore, we investigate the robustness of QIRO and RQAOA against perturbations of the optimal parameters of the QAOA circuits generating the correlations.
This allows us to emulate failures of the classical optimizer to find the optimal parameters.
Notably, we also deploy QIRO on the QuEra Aquila neutral atom device (as available on Amazon Braket), to (approximately) solve the MIS problem on unit disk graphs with more than a hundred nodes.

\subsection{MIS}
\label{sec:mis-results}

\begin{figure}
\includegraphics{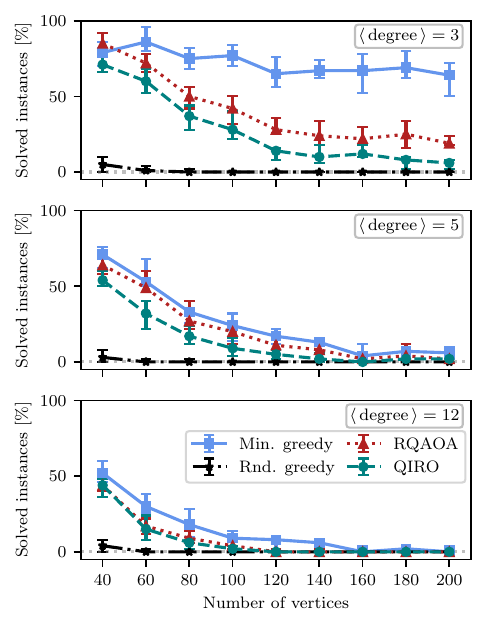}
\caption{System size scaling for the success rate of the random greedy (Rnd. greedy), minimal degree greedy (Min. greedy), RQAOA, and QIRO for Erd\H{o}s-R\'enyi graphs with different expectation values of the average degree. At each system size and average degree fifty graphs were generated, and ten runs of each algorithm were performed. We report the median percentage of graphs for which we find independent sets of the size found by the classical \texttt{ReduMis} benchmark algorithm. The error bars denote the minimum and maximum performance observed across the ten runs. Note that the results for QIRO and RQAOA are obtained using simulations of QAOA at depth $p=1$.
}
\label{fig:mis-scaling}
\end{figure}

We begin by presenting the results of QIRO and RQAOA as applied to the MIS problem. First, we consider the algorithmic performance as a function of the system size on the well-studied family of Erd\H{o}s-R\'enyi graphs, with fixed average degrees~\cite{Erdos2022-hw}. As a classical comparison, we use two variants of a greedy algorithm. This choice is motivated by recent findings that information from quantum devices can be used to enhance the performance of greedy solvers~\cite{dupont_quantum_2023}. First, we consider a simple \emph{random greedy} algorithm, which iteratively assigns random nodes to the independent set, and removes nodes adjacent to the selected node such that no violations are possible. As the second comparison we use the \emph{minimal degree greedy} algorithm, which iteratively assigns a node with the minimal degree to the independent set and likewise removes adjacent nodes to ensure the validity of the final solution~\cite{halldorsson_greed_1997}. We note that the minimal degree greedy algorithm was designed chiefly for sparse graphs (i.e., of low average degree). This is indicated by the performance guarantee of the greedy algorithm, which is inversely proportional to the graph's average degree~\cite{halldorsson_greed_1997}. However, we note that even for the sparse graphs considered throughout this section, stronger algorithms exist in the literature -- see Ref.~\cite{boether} for benchmarks.  

Importantly, because all of the algorithms considered here operate by iteratively simplifying the problem, their performance is indicative of the fitness of the information used to perform these simplifications.

For each system size and average degree we generated a set of fifty random Erd\H{o}s-R\'enyi graphs and solved these with the \texttt{ReduMis} heuristic solver from the \texttt{KaMIS} project~\cite{lamm_finding_2017, hespe_scalable_2019}. We use the solution obtained by \texttt{ReduMis} as (a proxy for) the optimal solution, and report the percentage of graphs for which each of the algorithms considered performs on par with the \texttt{ReduMis} reference.

\begin{figure*}[ht]
\centering
\subfloat[]{%
\includegraphics[width=5.8cm]{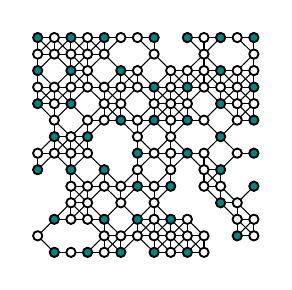}%
\label{fig:example-graph}%
}
\subfloat[]{%
\includegraphics{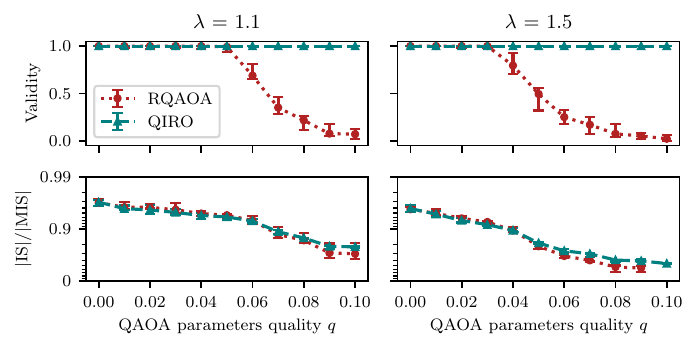}%
\label{fig:mis-validity-apprat}
}
\caption{
(a) An example unit disk graph  with 137 nodes. Teal nodes indicate a maximum independent set of size $|\mathrm{MIS}|=45$.  (b) Data for 500 random unit disk graphs with 137 nodes, comparing the robustness of RQAOA and QIRO, for two different magnitudes of the penalty term $\lambda$ (see Eq.~\eqref{eq:mis-hamiltonian}). The upper row shows the fraction of valid solutions obtained by each algorithm. The lower row displays the ratio between the size of the largest independent set (IS) found and the maximum independent set (MIS) size. Both the validity and $|\mathrm{IS}|/\mathrm{MIS}|$ are plotted with respect to the quality of the $p=1$ QAOA parameters, with parameters at $q=0.0$ corresponding to the optimal parameters, and higher values of $q$ corresponding to worse parameters (see main text). We note that we only report the approximation ratio achieved by RQAOA if the solution produced by it is valid (i.e., corresponds to an independent set).
}
\end{figure*}

The results for graph sizes with $n=40$ -- $200$ nodes and average degrees of $3$, $5$, and $12$ are shown in Fig.~\ref{fig:mis-scaling}. We used $\lambda = 1.1$ in the MIS Hamiltonian (see Eq.~\eqref{eq:mis-hamiltonian}) for RQAOA and QIRO. The performance of the two quantum algorithms (QIRO and RQAOA) is found to lie between the random and minimal degree greedy algorithms. This highlights the limitations of the quantum correlations obtained from QAOA at $p=1$, which only ``sees'' the immediate neighbors of every node. As such, it is unsurprising that the minimal degree greedy algorithm ---which makes the optimal greedy move based on information from nearest neighbors---outperforms QIRO and RQAOA informed by $p=1$ QAOA.
However, our results in Appendix~\ref{app:higher-depth-qaoa} indicate that (at least on small graphs), QIRO with QAOA at $p>1$ outperforms the greedy benchmark.

Furthermore, RQAOA (slightly, but consistently) outperforms QIRO. This is likely due to the fact that the QIRO simplification rules as defined in Sec.~\ref{sec:mis-update-rules} are manifestly local. The fact that the gap between QIRO and RQAOA seems to diminish for denser graphs (i.e., for increasing average degree) corroborates this intuitive explanation, as we expect locality to be a weaker limitation if the underlying interaction graph has a higher connectivity.

A major advantage of QIRO (as compared to, e.g., RQAOA) is that the problem-specific classical subroutines can be used to enforce feasibility of the solutions generated (i.e., all obtained candidate solutions are guaranteed to be independent sets). To showcase this feature, we performed numerical simulations of QIRO and RQAOA with $p=1$ QAOA with suboptimal parameters. The goal of these experiments is to consider the robustness of the algorithms against lower-quality quantum information. The parameters are generated by sampling a uniform $30\times 30$ grid in the $(\beta, \gamma)$-parameter plane, and choosing the parameters corresponding to $q$-th quantile in the energy values, among the $30\times 30$ energies obtained. Given their relevance for neutral atom quantum devices~\cite{pichler_quantum_2018, ebadi_quantum_2022}, we use unit disk graphs (UDGs) (see Appendix~\ref{app:quera-details}) for these experiments.

The results for an ensemble of fifty UDGs with 137 nodes and two different values of the penalty term $\lambda$ are shown in Fig.~\ref{fig:mis-validity-apprat}. For every graph and parameter quality $q$ ten runs of each algorithm were performed. We show the mean approximation ratio of valid solutions produced by RQAOA and QIRO as plotted against the quality quantile $q$ of the parameters used to generate the correlations by the QAOA. The parameters at $q=0$ correspond to the fully optimized parameters. As expected, QIRO only produces valid solutions. Remarkably, if perfect (i.e., optimized) parameters are supplied to RQAOA, it likewise produces exclusively valid solutions. However, for both values of $\lambda$, there is a sharp drop-off in the number of valid solutions for RQAOA. Because finding optimal parameters in an actual experimental setting is challenging ~\cite{lee_progress_2021,Bittel_2021}, the robustness of QIRO is a valuable feature, in particular for realistic experimental scenarios. Additionally, results in Figure~\ref{fig:mis-validity-apprat} suggest that increasing the quality of the quantum correlations has a positive impact on the performance of both QIRO as well as RQAOA. This is in agreement with the outcome of our experiments at $p>1$ in Appendix~\ref{app:higher-depth-qaoa}, where a clear of improved QIRO performance with increasing $p$ is observed.


Surprisingly, results shown in Fig.~\ref{fig:mis-validity-apprat} indicate that the number of valid solutions produced by RQAOA falls off at a comparatively better parameter quality (lower $q$) for the larger value of $\lambda$, defying the expectations that a larger penalty term $\lambda$ should favor valid solutions. In general, we observe complex behaviour in the $(\lambda, q)$ parameter plane (see Fig.~\ref{fig:validity-matrix} and the associated discussion in Appendix~\ref{app:rqaoa-validity}). We note that determining the magnitude of penalty terms is a well-known issue when incorporating hard constraints into an unconstrained optimization setup~\cite{lucas_ising_2014, Glover2018}.

Importantly, at each step of the QIRO algorithm, what remains to be solved is a valid MIS problem. Moreover, if the starting graph is a UDG (as in Fig.~\ref{fig:example-graph}) then also all successive (simplified) graphs are -- see also Fig.~\ref{fig:mis-reductions}. Because UDGs can be naturally embedded on neutral atom quantum processors~\cite{pichler_quantum_2018, ebadi_quantum_2022}, we can use such devices to produce the required quantum correlations. In what follows, we present results from the QuEra Aquila neutral atom quantum processor, as accessed through Amazon Braket. The details on the (quasi-)adiabatic protocols used to generate the correlations on the QuEra Aquila device can be found in Appendix~\ref{app:quera-details}. Due to the limited accessibility of the device, we report the results on a per-instance basis. We sort the instances in increasing order of their classical hardness [see Eq.~\eqref{eq:hardness-parameter}] as proposed by Ebadi \emph{et al}.~\cite{ebadi_quantum_2022}. We provide details about the hardness of the considered graph instances in Appendix~\ref{app:quera-details}.

In Fig.~\ref{fig:quera-data} we display the approximation ratios $|\mathrm{IS}|/|\mathrm{MIS}|$ obtained by QIRO with correlations from the QuEra device (QIRO-QuEra) and classical simulations of $p=1$ QAOA (QIRO-QAOA), respectively. On the QuEra device, we generate the correlations using the protocols outlined in Appendix~\ref{app:quera-details}. We compare these to the sizes of the independent sets found by the minimal degree greedy algorithm. The reference maximum independent set is computed using the tensor network algorithm from Ref.~\cite{liu_computing_2023}. For clarity, results of RQAOA are omitted from the plot, because the approximation ratios found for UDGs were generally comparable to those found by QIRO-QAOA (as seen in Fig.~\ref{fig:mis-validity-apprat}).

\begin{figure*}
\centering
\includegraphics{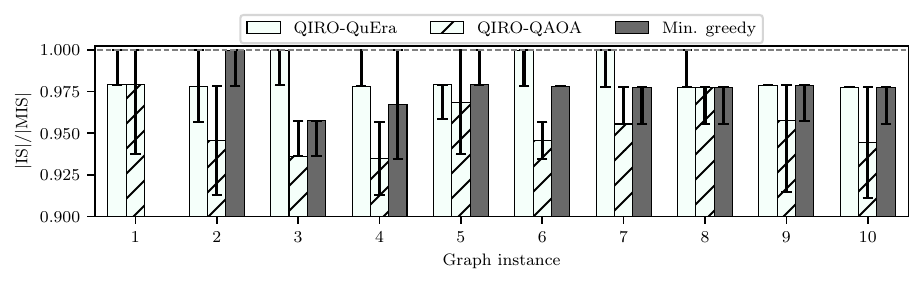}
\caption{
Approximation ratio comparing the size of the independent set $|\mathrm{IS}|$ found by QIRO with correlations from the QuEra neutral atom quantum processor (QIRO-QuEra), and correlations from numerical simulations of $p=1$ QAOA (QIRO-QAOA), with the size of the maximum independent set $|\mathrm{MIS}|$. For comparison, the ratios obtained by the minimal degree greedy algorithm (Min.~greedy) are shown. In instance $1$ the obtained approximation ratio of the greedy algorithm is outside of the range of the displayed values with $|\mathrm{IS}|/|\mathrm{MIS}|\approx 0.63$. The median value over five runs is plotted, with error bars showing the best and worst solutions found.
}
\label{fig:quera-data}%
\end{figure*}

Our results indicate that the quantum correlations from the QuEra machine are better at guiding the QIRO procedure than numerical simulations of QAOA at $p=1$. This is in line with previous results for solving the MIS problem using analog protocols on such hardware~\cite{ebadi_quantum_2022,finzgar_bo_2023}. A larger sample size would be required to solidify such claims. However, initial results appear to suggest that QIRO-QuEra performs comparably to the minimal degree greedy algorithm, reinforcing our belief that improved quantum correlations boost the performance of QIRO. Strikingly, the minimal degree greedy algorithm fails dramatically on some graphs (e.g., instance 1), where it achieves an approximation ratio of around $\sim 0.6$, while QIRO-QuEra is able to solve the same problems (nearly) optimally. This preliminary evidence showcases the potential utility of quantum correlations for guiding optimization algorithms.

\subsection{MAX-2-SAT}
\label{sec:max2sat-results}

Next, we analyze the performance of QIRO and QIRO with backtracking (QIRO + BT) on random MAX-2-SAT instances with clauses of length two. We compare our results to those obtained with RQAOA and simulated annealing (SA)~\cite{kirkpatrick-sa}, an established and relatively simple classical heuristic. More detailed simulations with other, more powerful heuristics are left for future studies. Details about our implementation of SA can be found in Appendix~\ref{app:pt-sa-setup}. 

We first analyze the system size scaling of the algorithms' performance for random MAX-2-SAT instances with up to $160$ variables, at three different values of the clause-to-variable ratio $\alpha$. We use the MAX-SAT solver RC2~\cite{Ignatiev2019} to compute the optimal solutions, and use parallel tempering (PT)~\cite{Hukushima_1996} to compute (approximately) optimal solutions for instances where the RC2 solver timed-out. Details of our implementation of PT can be found in Appendix~\ref{app:pt-sa-setup}. We then report the number of times the algorithms perform at least as well as PT or the RC2 solver.

\begin{figure}
\includegraphics{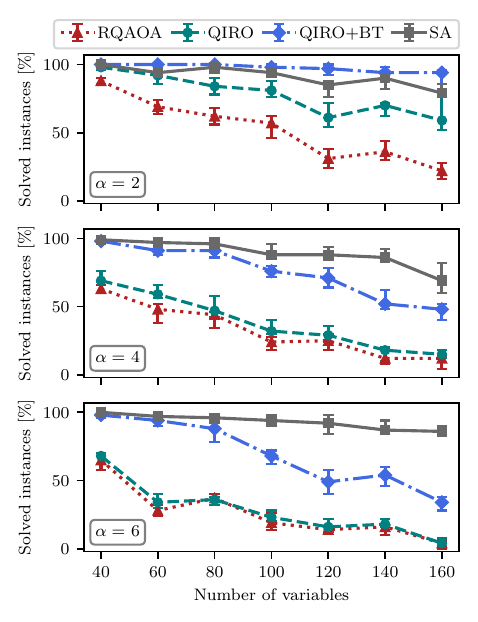}
\caption{System size scaling of the performance of the different algorithms, at three different values of $\alpha$. At each system size and $\alpha$, fifty random instances were generated with 2 variables per clause, each variable and its polarity chosen uniformly at random. Parallel tempering (PT) was used to find ${\text{(nearly-)optimal}}$ solutions; we then report the median percentage of instances on which an algorithm performs at least as well as PT across 10 independent runs. The error bars indicate the minimal and maximal performance across these ten runs. Simulation details about SA and PT can be found in Appendix~\ref{app:pt-sa-setup}.
}
\label{fig:max-sat-over-variables}
\end{figure}

Figure~\ref{fig:max-sat-over-variables} showcases the results of our experiments. QIRO performs better than RQAOA for all clause-to-variable ratios $\alpha$ and over the entire range of the considered problem sizes. This shows that performance gains can be made by including very simple problem-specific update rules that require only modest classical resources. We note, however, that the improvement is especially prominent for lower values of $\alpha$, and diminishes for problems with higher $\alpha$ (connectivity). Moreover, due to the inclusion of inference rules, the Boolean formula is simplified in fewer update steps, thus requiring fewer calls to the quantum device (and fewer optimizations of the QAOA circuit) than in RQAOA.

As expected, we notice a decrease in performance with increasing problem sizes, with both QIRO and RQAOA often struggling to find optimal solutions at the largest problem sizes considered. However, with the addition of backtracking, QIRO+BT is able to optimally solve more than half of the instances for almost all system sizes and all values of $\alpha$ considered here. For $\alpha = 2$ it even slightly outperforms our SA implementation. It should be noted that while the number of calls to the quantum device increases significantly (i.e., from linear to quadratic in the problem size) if backtracking is included, the requirements on the quantum hardware in terms of the gate and qubit count are kept constant. This fundamentally differs from QAOA, where an improved performance can only be obtained by increasing the circuit depth $p$, and thus the quantum resources required.

\section{Discussion}
\label{sec:discussion}

In this work we have extended previous efforts on hybrid quantum-classical algorithms for combinatorial optimization, by introducing a family of quantum-informed recursive optimization (QIRO) algorithms. We propose the use of quantum-informed classical update steps to recursively simplify the original optimization problem, allowing for the incorporation of problem-specific classical subroutines (e.g., satisfiability inference rules) from the rich literature on classical combinatorial optimization~\cite{Papadimitriou1998-pc}. We have shown via numerical simulations for random instances of two NP-hard combinatorial optimization problems that such classical subroutines can significantly strengthen the algorithm's performance. Additionally, we compare the performance of QIRO to RQAOA, providing, to the best of our knowledge, the first systematic numerical study of RQAOA on problems that are not $\mathds{Z}_2$-symmetric.

Importantly, we think that QIRO offers a promising approach for solving optimization problems with \emph{hard} constraints, which have previously proven challenging for quantum algorithms~\cite{lucas_ising_2014, Glover2018, hadfield_quantum_2019}.  Using the MIS problem as a guiding example, we have shown how classical update steps can be used to enforce hard constraints and only produce feasible solutions. This is true regardless of the quality of the quantum resources, endowing QIRO with the robustness essential for algorithms within the NISQ era. 

Interestingly, we observe that keeping the quantum resources equal, RQAOA outperforms QIRO on the MIS problem. We believe that this is a consequence of the local nature of the QIRO update rules for MIS. This is corroborated by our finding that improved quantum correlations from the neutral atom device enhance QIRO's performance compared to when QAOA at $p=1$ is used to generate the correlations. The intuitive explanation is that the quantum many-body evolution on a neutral atom quantum processor provides sufficient non-local effects to counteract the locality of the update steps. In fact, adiabatic protocols analogous to those used to generate correlations here have previously been shown to produce candidate solutions surpassing those produced by QAOA at depths higher than $p=1$~\cite{ebadi_quantum_2022}. Taken together, our results may serve as yet another indication that locality is a limiting factor for (quantum) optimization algorithms~\cite{Hastings2019}.

Furthermore, we have confirmed the intuitive expectation that higher-quality quantum correlations should enhance the performance of the algorithm, as evidenced by means of extensive numerical simulations, as well as experiments on an actual neutral atom quantum device. This result suggests that advances in quantum hardware should lead to further performance gains.

We note that we performed no optimization of the protocols used to generate the correlations in our experiments on the QuEra Aquila neutral atom device. Previously, dramatic improvements in the quality of candidate solutions has been obtained by first optimizing the schedules~\cite{ebadi_quantum_2022, finzgar_bo_2023}. Moreover, the (unoptimized) protocols and device used in this study (and specified in Appendix~\ref{app:quera-details}) have been shown to produce candidate solutions typically corresponding to either very small independent sets, or possessing multiple violations of the independence constraint~\cite{finzgar_bo_2023}. Thus, it is reasonable to expect that such schedule optimization schemes could greatly benefit the quantum state preparation step in QIRO. Moreover, the ability of QIRO to find good (feasible) solutions despite the limitations of the quantum hardware serves as yet another demonstration of its robustness.

In addition, we have introduced the use of backtracking to obtain improved results via additional calls to existing quantum hardware, rather than requiring deeper circuits, as is usually the case in quantum (optimization) algorithms. This scheme has led to promising results for the MAX-2-SAT problem, where QIRO enhanced with backtracking could optimally solve most random problem instances, even at comparatively large instance sizes. Because large instances are accessible in numerical simulations (at $p=1$ of QAOA), we were able to sensibly compare the performance of QIRO(+BT) with widely-used classical heuristic solvers such as simulated annealing. 

Finally, we would like to stress that our work should be seen as a first demonstration of a general template for a larger class of hybrid quantum-classical algorithms for combinatorial optimization. We firmly believe that hybrid techniques will be necessary to unlock the advantages of quantum devices. The modular nature of QIRO offers a plethora of ways in which the presented toolbox can be modified; we touch upon some potential extensions in the next section.
    
\section{Outlook}
\label{sec:outlook}

Several modifications to the QIRO algorithm as presented here can be envisioned. Exploring update rules for optimization problems beyond those examined in this manuscript would be a compelling avenue for future research. Likewise, it would be intriguing to devise the update rules for weighted versions of the problems addressed here. Moreover, one could attempt to extend the information extracted from the quantum state beyond just the maximal correlation. This could be accomplished by a randomized strategy, with the selection probabilities determined as some function of the correlation matrix. Furthermore, devising update rules incorporating correlations beyond those between nearest neighbors in the interaction graph might be useful, especially in combination with deeper QAOA circuits or on analog devices with long-range interactions (e.g., Rydberg atom arrays). Additionally, one could exploit the locality of QAOA, and simultaneously round multiple correlations at each update step, if the corresponding variables are outside of each other's sphere of influence. Thus, one could save resources by requiring fewer calls to the quantum routines, and thus mitigate one of the limitations of the presented scheme. On the other hand, if deeper quantum circuits (e.g., QAOA at $p>1$) are available, extracting correlations beyond those between variables adjacent in the interaction graph might be beneficial.
Finally, considering problem-tailored alternatives to a brute-force search of the remaining solution space at a threshold problem size $n_{\mathbf{c}}$ may significantly improve the performance and resource-efficiency of QIRO.

It is unreasonable to expect that the simple backtracking strategy presented here would be equally successful for larger problem instances, or instances from different problem classes. Hence, one might need to devise improved backtracking strategies, both in terms of the resources required, as well as in terms of the performance. One possibility is the use of reinforcement learning to guide the reduction process, as proposed by Patel~\emph{et. al}~\cite{patel_reinforcement_2022_1}. Resources might be saved by efficient pruning of the search tree using existing methods~\cite{Abrame2015}, identifying and ignoring unfavorable parts of the solution space. Therefore, fewer quantum state preparations would be required. Finally, one could use backtracking to enforce constraints, by identifying steps where a constraint was violated, backtracking to it, and altering the decision made at that step.

It would likewise be interesting to consider alternative methods of generating the quantum correlations. This could include efficient numerical methods to obtain correlations with the QAOA at $p>1$, by means of, e.g., tensor network simulations~\cite{Huang2021, Gray_2021}. Naturally, it would be interesting to acquire correlations from higher depth QAOA circuits executed on real digital quantum computers, or by generating them using quantum annealers based on superconducting flux qubits~\cite{Johnson_2011,King_2022}. One could consider optimizing analog protocols (using e.g., Bayesian optimization~\cite{finzgar_bo_2023}) to obtain better correlations, and enhance the effectiveness of the algorithm. Moreover, modifying the cost function with respect to which the quantum resources are optimized might be helpful, similarly to the approach employed by Caha~\emph{et al.}~\cite{caha_twisted_2022}. Finally, one could attempt to generate correlations using classical approximation algorithms, as proposed by Wagner~\emph{et al.}~\cite{wagner2023enhancing}. Long story short, there is a lot to be done, but this paper is already long enough. 

Conveniently, the modular nature of QIRO allows us to analyze the promise of quantum computers for combinatorial optimization, e.g., by comparing QIRO with different quantum and classical methods of computing the correlations. Should shortcomings of quantum approaches (e.g., locality) be identified in this manner, the classical subroutines in QIRO can be designed to rectify them. As such, we hope that our work can bring us closer to determining whether quantum devices can provide value for combinatorial optimization problems.

\section*{Acknowledgements}
JRF and AK thank Libor Caha and Alexander Kliesch for insightful discussions. The authors thank Lilly Palackal, Maximilian Passek, Carlos Riofr\'{\i}o and Gili Rosenberg for thorough reviews of the manuscript, and the Amazon Braket, BMW, and QuEra teams for their support. CM thanks the Munich Quantum Valley initiative, which is supported by the Bavarian state government with funds from the Hightech Agenda Bayern Plus. HGK would like to thank Am Platzl 1A for providing the necessary environment for creative thinking. An open-source implementation of QIRO is available online~\footnote{\href{https://github.com/jernejrudifinzgar/qiro}{https://github.com/jernejrudifinzgar/qiro}}.

%


\begin{thebibliography}{67}%
\makeatletter
\providecommand \@ifxundefined [1]{%
 \@ifx{#1\undefined}
}%
\providecommand \@ifnum [1]{%
 \ifnum #1\expandafter \@firstoftwo
 \else \expandafter \@secondoftwo
 \fi
}%
\providecommand \@ifx [1]{%
 \ifx #1\expandafter \@firstoftwo
 \else \expandafter \@secondoftwo
 \fi
}%
\providecommand \natexlab [1]{#1}%
\providecommand \enquote  [1]{``#1''}%
\providecommand \bibnamefont  [1]{#1}%
\providecommand \bibfnamefont [1]{#1}%
\providecommand \citenamefont [1]{#1}%
\providecommand \href@noop [0]{\@secondoftwo}%
\providecommand \href [0]{\begingroup \@sanitize@url \@href}%
\providecommand \@href[1]{\@@startlink{#1}\@@href}%
\providecommand \@@href[1]{\endgroup#1\@@endlink}%
\providecommand \@sanitize@url [0]{\catcode `\\12\catcode `\$12\catcode
  `\&12\catcode `\#12\catcode `\^12\catcode `\_12\catcode `\%12\relax}%
\providecommand \@@startlink[1]{}%
\providecommand \@@endlink[0]{}%
\providecommand \url  [0]{\begingroup\@sanitize@url \@url }%
\providecommand \@url [1]{\endgroup\@href {#1}{\urlprefix }}%
\providecommand \urlprefix  [0]{URL }%
\providecommand \Eprint [0]{\href }%
\providecommand \doibase [0]{http://dx.doi.org/}%
\providecommand \selectlanguage [0]{\@gobble}%
\providecommand \bibinfo  [0]{\@secondoftwo}%
\providecommand \bibfield  [0]{\@secondoftwo}%
\providecommand \translation [1]{[#1]}%
\providecommand \BibitemOpen [0]{}%
\providecommand \bibitemStop [0]{}%
\providecommand \bibitemNoStop [0]{.\EOS\space}%
\providecommand \EOS [0]{\spacefactor3000\relax}%
\providecommand \BibitemShut  [1]{\csname bibitem#1\endcsname}%
\let\auto@bib@innerbib\@empty
\bibitem [{\citenamefont {Farhi}\ \emph {et~al.}(2014)\citenamefont {Farhi},
  \citenamefont {Goldstone},\ and\ \citenamefont {Gutmann}}]{Farhi2014-qaoa}%
  \BibitemOpen
  \bibfield  {author} {\bibinfo {author} {\bibfnamefont {E.}~\bibnamefont
  {Farhi}}, \bibinfo {author} {\bibfnamefont {J.}~\bibnamefont {Goldstone}}, \
  and\ \bibinfo {author} {\bibfnamefont {S.}~\bibnamefont {Gutmann}},\
  }\href@noop {} {\  (\bibinfo {year} {2014})},\ \Eprint
  {http://arxiv.org/abs/1411.4028} {arXiv:1411.4028} \BibitemShut {NoStop}%
\bibitem [{\citenamefont {Gamarnik}(2021)}]{gamarnik_overlap_2021}%
  \BibitemOpen
  \bibfield  {author} {\bibinfo {author} {\bibfnamefont {D.}~\bibnamefont
  {Gamarnik}},\ }\href {\doibase 10.1073/pnas.2108492118} {\bibfield  {journal}
  {\bibinfo  {journal} {Proceedings of the National Academy of Sciences}\
  }\textbf {\bibinfo {volume} {118}},\ \bibinfo {pages} {e2108492118} (\bibinfo
  {year} {2021})}\BibitemShut {NoStop}%
\bibitem [{\citenamefont {Bravyi}\ \emph {et~al.}(2020)\citenamefont {Bravyi},
  \citenamefont {Kliesch}, \citenamefont {Koenig},\ and\ \citenamefont
  {Tang}}]{bravyi_obstacles_2020}%
  \BibitemOpen
  \bibfield  {author} {\bibinfo {author} {\bibfnamefont {S.}~\bibnamefont
  {Bravyi}}, \bibinfo {author} {\bibfnamefont {A.}~\bibnamefont {Kliesch}},
  \bibinfo {author} {\bibfnamefont {R.}~\bibnamefont {Koenig}}, \ and\ \bibinfo
  {author} {\bibfnamefont {E.}~\bibnamefont {Tang}},\ }\href {\doibase
  10.1103/PhysRevLett.125.260505} {\bibfield  {journal} {\bibinfo  {journal}
  {Physical Review Letters}\ }\textbf {\bibinfo {volume} {125}},\ \bibinfo
  {pages} {260505} (\bibinfo {year} {2020})}\BibitemShut {NoStop}%
\bibitem [{\citenamefont {Farhi}\ \emph
  {et~al.}(2020{\natexlab{a}})\citenamefont {Farhi}, \citenamefont {Gamarnik},\
  and\ \citenamefont {Gutmann}}]{farhi_quantum_2020-1}%
  \BibitemOpen
  \bibfield  {author} {\bibinfo {author} {\bibfnamefont {E.}~\bibnamefont
  {Farhi}}, \bibinfo {author} {\bibfnamefont {D.}~\bibnamefont {Gamarnik}}, \
  and\ \bibinfo {author} {\bibfnamefont {S.}~\bibnamefont {Gutmann}},\
  }\href@noop {} {\  (\bibinfo {year} {2020}{\natexlab{a}})},\ \Eprint
  {http://arxiv.org/abs/2004.09002} {arXiv:2004.09002} \BibitemShut {NoStop}%
\bibitem [{\citenamefont {Farhi}\ \emph
  {et~al.}(2020{\natexlab{b}})\citenamefont {Farhi}, \citenamefont {Gamarnik},\
  and\ \citenamefont {Gutmann}}]{farhi_quantum_2020-2}%
  \BibitemOpen
  \bibfield  {author} {\bibinfo {author} {\bibfnamefont {E.}~\bibnamefont
  {Farhi}}, \bibinfo {author} {\bibfnamefont {D.}~\bibnamefont {Gamarnik}}, \
  and\ \bibinfo {author} {\bibfnamefont {S.}~\bibnamefont {Gutmann}},\
  }\href@noop {} {\  (\bibinfo {year} {2020}{\natexlab{b}})},\ \Eprint
  {http://arxiv.org/abs/2005.08747} {arXiv:2005.08747} \BibitemShut {NoStop}%
\bibitem [{\citenamefont {Chou}\ \emph {et~al.}(2022)\citenamefont {Chou},
  \citenamefont {Love}, \citenamefont {Sandhu},\ and\ \citenamefont
  {Shi}}]{Chou2021}%
  \BibitemOpen
  \bibfield  {author} {\bibinfo {author} {\bibfnamefont {C.-N.}\ \bibnamefont
  {Chou}}, \bibinfo {author} {\bibfnamefont {P.~J.}\ \bibnamefont {Love}},
  \bibinfo {author} {\bibfnamefont {J.~S.}\ \bibnamefont {Sandhu}}, \ and\
  \bibinfo {author} {\bibfnamefont {J.}~\bibnamefont {Shi}},\ }in\ \href
  {\doibase 10.4230/LIPIcs.ICALP.2022.41} {\emph {\bibinfo {booktitle} {49th
  International Colloquium on Automata, Languages, and Programming (ICALP
  2022)}}},\ \bibinfo {series} {Leibniz International Proceedings in
  Informatics (LIPIcs)}, Vol.\ \bibinfo {volume} {229},\ \bibinfo {editor}
  {edited by\ \bibinfo {editor} {\bibfnamefont {M.}~\bibnamefont
  {Boja\'{n}czyk}}, \bibinfo {editor} {\bibfnamefont {E.}~\bibnamefont
  {Merelli}}, \ and\ \bibinfo {editor} {\bibfnamefont {D.~P.}\ \bibnamefont
  {Woodruff}}}\ (\bibinfo  {publisher} {Schloss Dagstuhl -- Leibniz-Zentrum
  f{\"u}r Informatik},\ \bibinfo {address} {Dagstuhl, Germany},\ \bibinfo
  {year} {2022})\ pp.\ \bibinfo {pages} {41:1--41:20}\BibitemShut {NoStop}%
\bibitem [{\citenamefont {Bravyi}\ \emph {et~al.}(2022)\citenamefont {Bravyi},
  \citenamefont {Kliesch}, \citenamefont {Koenig},\ and\ \citenamefont
  {Tang}}]{bravyi_hybrid_2022}%
  \BibitemOpen
  \bibfield  {author} {\bibinfo {author} {\bibfnamefont {S.}~\bibnamefont
  {Bravyi}}, \bibinfo {author} {\bibfnamefont {A.}~\bibnamefont {Kliesch}},
  \bibinfo {author} {\bibfnamefont {R.}~\bibnamefont {Koenig}}, \ and\ \bibinfo
  {author} {\bibfnamefont {E.}~\bibnamefont {Tang}},\ }\href {\doibase
  10.22331/q-2022-03-30-678} {\bibfield  {journal} {\bibinfo  {journal}
  {Quantum}\ }\textbf {\bibinfo {volume} {6}},\ \bibinfo {pages} {678}
  (\bibinfo {year} {2022})}\BibitemShut {NoStop}%
\bibitem [{\citenamefont {McClean}\ \emph {et~al.}(2021)\citenamefont
  {McClean}, \citenamefont {Harrigan}, \citenamefont {Mohseni}, \citenamefont
  {Rubin}, \citenamefont {Jiang}, \citenamefont {Boixo}, \citenamefont
  {Smelyanskiy}, \citenamefont {Babbush},\ and\ \citenamefont
  {Neven}}]{mcclean_low-depth_2021}%
  \BibitemOpen
  \bibfield  {author} {\bibinfo {author} {\bibfnamefont {J.~R.}\ \bibnamefont
  {McClean}}, \bibinfo {author} {\bibfnamefont {M.~P.}\ \bibnamefont
  {Harrigan}}, \bibinfo {author} {\bibfnamefont {M.}~\bibnamefont {Mohseni}},
  \bibinfo {author} {\bibfnamefont {N.~C.}\ \bibnamefont {Rubin}}, \bibinfo
  {author} {\bibfnamefont {Z.}~\bibnamefont {Jiang}}, \bibinfo {author}
  {\bibfnamefont {S.}~\bibnamefont {Boixo}}, \bibinfo {author} {\bibfnamefont
  {V.~N.}\ \bibnamefont {Smelyanskiy}}, \bibinfo {author} {\bibfnamefont
  {R.}~\bibnamefont {Babbush}}, \ and\ \bibinfo {author} {\bibfnamefont
  {H.}~\bibnamefont {Neven}},\ }\href {\doibase 10.1103/PRXQuantum.2.030312}
  {\bibfield  {journal} {\bibinfo  {journal} {PRX Quantum}\ }\textbf {\bibinfo
  {volume} {2}},\ \bibinfo {pages} {030312} (\bibinfo {year}
  {2021})}\BibitemShut {NoStop}%
\bibitem [{\citenamefont {Karimi}\ \emph {et~al.}(2017)\citenamefont {Karimi},
  \citenamefont {Rosenberg},\ and\ \citenamefont
  {Katzgraber}}]{karimi_effective_2017}%
  \BibitemOpen
  \bibfield  {author} {\bibinfo {author} {\bibfnamefont {H.}~\bibnamefont
  {Karimi}}, \bibinfo {author} {\bibfnamefont {G.}~\bibnamefont {Rosenberg}}, \
  and\ \bibinfo {author} {\bibfnamefont {H.~G.}\ \bibnamefont {Katzgraber}},\
  }\href {\doibase 10.1103/PhysRevE.96.043312} {\bibfield  {journal} {\bibinfo
  {journal} {Physical Review E}\ }\textbf {\bibinfo {volume} {96}},\ \bibinfo
  {pages} {043312} (\bibinfo {year} {2017})}\BibitemShut {NoStop}%
\bibitem [{\citenamefont {Lucas}(2014)}]{lucas_ising_2014}%
  \BibitemOpen
  \bibfield  {author} {\bibinfo {author} {\bibfnamefont {A.}~\bibnamefont
  {Lucas}},\ }\href {\doibase 10.3389/fphy.2014.00005} {\bibfield  {journal}
  {\bibinfo  {journal} {Frontiers in Physics}\ }\textbf {\bibinfo {volume} {2}}
  (\bibinfo {year} {2014}),\ 10.3389/fphy.2014.00005}\BibitemShut {NoStop}%
\bibitem [{\citenamefont {Glover}\ \emph {et~al.}(2019)\citenamefont {Glover},
  \citenamefont {Kochenberger},\ and\ \citenamefont {Du}}]{Glover2018}%
  \BibitemOpen
  \bibfield  {author} {\bibinfo {author} {\bibfnamefont {F.}~\bibnamefont
  {Glover}}, \bibinfo {author} {\bibfnamefont {G.}~\bibnamefont
  {Kochenberger}}, \ and\ \bibinfo {author} {\bibfnamefont {Y.}~\bibnamefont
  {Du}},\ }\href@noop {} {\  (\bibinfo {year} {2019})},\ \Eprint
  {http://arxiv.org/abs/1811.11538} {arXiv:1811.11538} \BibitemShut {NoStop}%
\bibitem [{\citenamefont {Butenko}\ and\ \citenamefont
  {Pardalos}(2003)}]{mis_app}%
  \BibitemOpen
  \bibfield  {author} {\bibinfo {author} {\bibfnamefont {S.}~\bibnamefont
  {Butenko}}\ and\ \bibinfo {author} {\bibfnamefont {P.~M.}\ \bibnamefont
  {Pardalos}},\ }\emph {\bibinfo {title} {Maximum Independent Set and Related
  Problems, with Applications}},\ \href@noop {} {Ph.D. thesis},\ \bibinfo
  {address} {USA} (\bibinfo {year} {2003}),\ \bibinfo {note}
  {aAI3120100}\BibitemShut {NoStop}%
\bibitem [{\citenamefont {Pichler}\ \emph {et~al.}(2018)\citenamefont
  {Pichler}, \citenamefont {Wang}, \citenamefont {Zhou}, \citenamefont {Choi},\
  and\ \citenamefont {Lukin}}]{pichler_quantum_2018}%
  \BibitemOpen
  \bibfield  {author} {\bibinfo {author} {\bibfnamefont {H.}~\bibnamefont
  {Pichler}}, \bibinfo {author} {\bibfnamefont {S.-T.}\ \bibnamefont {Wang}},
  \bibinfo {author} {\bibfnamefont {L.}~\bibnamefont {Zhou}}, \bibinfo {author}
  {\bibfnamefont {S.}~\bibnamefont {Choi}}, \ and\ \bibinfo {author}
  {\bibfnamefont {M.~D.}\ \bibnamefont {Lukin}},\ }\href@noop {} {\  (\bibinfo
  {year} {2018})},\ \Eprint {http://arxiv.org/abs/1808.10816}
  {arXiv:1808.10816} \BibitemShut {NoStop}%
\bibitem [{\citenamefont {Ebadi}\ \emph {et~al.}(2022)\citenamefont {Ebadi},
  \citenamefont {Keesling}, \citenamefont {Cain}, \citenamefont {Wang},
  \citenamefont {Levine}, \citenamefont {Bluvstein}, \citenamefont {Semeghini},
  \citenamefont {Omran}, \citenamefont {Liu}, \citenamefont {Samajdar},
  \citenamefont {Luo}, \citenamefont {Nash}, \citenamefont {Gao}, \citenamefont
  {Barak}, \citenamefont {Farhi}, \citenamefont {Sachdev}, \citenamefont
  {Gemelke}, \citenamefont {Zhou}, \citenamefont {Choi}, \citenamefont
  {Pichler}, \citenamefont {Wang}, \citenamefont {Greiner}, \citenamefont
  {Vuleti\'{c}},\ and\ \citenamefont {Lukin}}]{ebadi_quantum_2022}%
  \BibitemOpen
  \bibfield  {author} {\bibinfo {author} {\bibfnamefont {S.}~\bibnamefont
  {Ebadi}}, \bibinfo {author} {\bibfnamefont {A.}~\bibnamefont {Keesling}},
  \bibinfo {author} {\bibfnamefont {M.}~\bibnamefont {Cain}}, \bibinfo {author}
  {\bibfnamefont {T.~T.}\ \bibnamefont {Wang}}, \bibinfo {author}
  {\bibfnamefont {H.}~\bibnamefont {Levine}}, \bibinfo {author} {\bibfnamefont
  {D.}~\bibnamefont {Bluvstein}}, \bibinfo {author} {\bibfnamefont
  {G.}~\bibnamefont {Semeghini}}, \bibinfo {author} {\bibfnamefont
  {A.}~\bibnamefont {Omran}}, \bibinfo {author} {\bibfnamefont {J.-G.}\
  \bibnamefont {Liu}}, \bibinfo {author} {\bibfnamefont {R.}~\bibnamefont
  {Samajdar}}, \bibinfo {author} {\bibfnamefont {X.-Z.}\ \bibnamefont {Luo}},
  \bibinfo {author} {\bibfnamefont {B.}~\bibnamefont {Nash}}, \bibinfo {author}
  {\bibfnamefont {X.}~\bibnamefont {Gao}}, \bibinfo {author} {\bibfnamefont
  {B.}~\bibnamefont {Barak}}, \bibinfo {author} {\bibfnamefont
  {E.}~\bibnamefont {Farhi}}, \bibinfo {author} {\bibfnamefont
  {S.}~\bibnamefont {Sachdev}}, \bibinfo {author} {\bibfnamefont
  {N.}~\bibnamefont {Gemelke}}, \bibinfo {author} {\bibfnamefont
  {L.}~\bibnamefont {Zhou}}, \bibinfo {author} {\bibfnamefont {S.}~\bibnamefont
  {Choi}}, \bibinfo {author} {\bibfnamefont {H.}~\bibnamefont {Pichler}},
  \bibinfo {author} {\bibfnamefont {S.-T.}\ \bibnamefont {Wang}}, \bibinfo
  {author} {\bibfnamefont {M.}~\bibnamefont {Greiner}}, \bibinfo {author}
  {\bibfnamefont {V.}~\bibnamefont {Vuleti\'{c}}}, \ and\ \bibinfo {author}
  {\bibfnamefont {M.~D.}\ \bibnamefont {Lukin}},\ }\href {\doibase
  10.1126/science.abo6587} {\bibfield  {journal} {\bibinfo  {journal}
  {Science}\ }\textbf {\bibinfo {volume} {376}},\ \bibinfo {pages} {1209}
  (\bibinfo {year} {2022})}\BibitemShut {NoStop}%
\bibitem [{\citenamefont {Wurtz}\ \emph {et~al.}(2022)\citenamefont {Wurtz},
  \citenamefont {Lopes}, \citenamefont {Gemelke}, \citenamefont {Keesling},\
  and\ \citenamefont {Wang}}]{wurtz_industry_2022_1}%
  \BibitemOpen
  \bibfield  {author} {\bibinfo {author} {\bibfnamefont {J.}~\bibnamefont
  {Wurtz}}, \bibinfo {author} {\bibfnamefont {P.}~\bibnamefont {Lopes}},
  \bibinfo {author} {\bibfnamefont {N.}~\bibnamefont {Gemelke}}, \bibinfo
  {author} {\bibfnamefont {A.}~\bibnamefont {Keesling}}, \ and\ \bibinfo
  {author} {\bibfnamefont {S.}~\bibnamefont {Wang}},\ }\href
  {http://arxiv.org/abs/2205.08500} {\  (\bibinfo {year} {2022})},\ \bibinfo
  {note} {arXiv:2205.08500 [quant-ph]}\BibitemShut {NoStop}%
\bibitem [{\citenamefont {Cook}(1971)}]{cook_1971}%
  \BibitemOpen
  \bibfield  {author} {\bibinfo {author} {\bibfnamefont {S.~A.}\ \bibnamefont
  {Cook}},\ }in\ \href {\doibase 10.1145/800157.805047} {\emph {\bibinfo
  {booktitle} {Proceedings of the third annual {ACM} symposium on {Theory} of
  computing - {STOC} '71}}}\ (\bibinfo  {publisher} {ACM Press},\ \bibinfo
  {address} {Shaker Heights, Ohio, United States},\ \bibinfo {year} {1971})\
  pp.\ \bibinfo {pages} {151--158}\BibitemShut {NoStop}%
\bibitem [{\citenamefont {Cheeseman}\ \emph {et~al.}(1991)\citenamefont
  {Cheeseman}, \citenamefont {Kanefsky},\ and\ \citenamefont
  {Taylor}}]{sat-phase-trans}%
  \BibitemOpen
  \bibfield  {author} {\bibinfo {author} {\bibfnamefont {P.}~\bibnamefont
  {Cheeseman}}, \bibinfo {author} {\bibfnamefont {B.}~\bibnamefont {Kanefsky}},
  \ and\ \bibinfo {author} {\bibfnamefont {W.~M.}\ \bibnamefont {Taylor}},\
  }in\ \href@noop {} {\emph {\bibinfo {booktitle} {Proceedings of the 12th
  International Joint Conference on Artificial Intelligence - Volume 1}}},\
  \bibinfo {series and number} {IJCAI'91}\ (\bibinfo  {publisher} {Morgan
  Kaufmann Publishers Inc.},\ \bibinfo {address} {San Francisco, CA, USA},\
  \bibinfo {year} {1991})\ p.\ \bibinfo {pages} {331–337}\BibitemShut
  {NoStop}%
\bibitem [{\citenamefont {Zeng}\ \emph {et~al.}(2016)\citenamefont {Zeng},
  \citenamefont {Zhang},\ and\ \citenamefont {Sarovar}}]{zeng_schedule_2016}%
  \BibitemOpen
  \bibfield  {author} {\bibinfo {author} {\bibfnamefont {L.}~\bibnamefont
  {Zeng}}, \bibinfo {author} {\bibfnamefont {J.}~\bibnamefont {Zhang}}, \ and\
  \bibinfo {author} {\bibfnamefont {M.}~\bibnamefont {Sarovar}},\ }\href
  {\doibase 10.1088/1751-8113/49/16/165305} {\bibfield  {journal} {\bibinfo
  {journal} {Journal of Physics A: Mathematical and Theoretical}\ }\textbf
  {\bibinfo {volume} {49}},\ \bibinfo {pages} {165305} (\bibinfo {year}
  {2016})}\BibitemShut {NoStop}%
\bibitem [{\citenamefont {Mezard}\ \emph {et~al.}(2002)\citenamefont {Mezard},
  \citenamefont {Parisi},\ and\ \citenamefont {Zecchina}}]{Mezard2002}%
  \BibitemOpen
  \bibfield  {author} {\bibinfo {author} {\bibfnamefont {M.}~\bibnamefont
  {Mezard}}, \bibinfo {author} {\bibfnamefont {G.}~\bibnamefont {Parisi}}, \
  and\ \bibinfo {author} {\bibfnamefont {R.}~\bibnamefont {Zecchina}},\ }\href
  {\doibase 10.1126/science.1073287} {\bibfield  {journal} {\bibinfo  {journal}
  {Science}\ }\textbf {\bibinfo {volume} {297}},\ \bibinfo {pages} {812}
  (\bibinfo {year} {2002})}\BibitemShut {NoStop}%
\bibitem [{\citenamefont {Sleegers}\ \emph {et~al.}(2020)\citenamefont
  {Sleegers}, \citenamefont {Olij}, \citenamefont {van Horn},\ and\
  \citenamefont {van~den Berg}}]{Sleegers2020}%
  \BibitemOpen
  \bibfield  {author} {\bibinfo {author} {\bibfnamefont {J.}~\bibnamefont
  {Sleegers}}, \bibinfo {author} {\bibfnamefont {R.}~\bibnamefont {Olij}},
  \bibinfo {author} {\bibfnamefont {G.}~\bibnamefont {van Horn}}, \ and\
  \bibinfo {author} {\bibfnamefont {D.}~\bibnamefont {van~den Berg}},\ }\href
  {\doibase 10.1016/j.orp.2020.100160} {\bibfield  {journal} {\bibinfo
  {journal} {Operations Research Perspectives}\ }\textbf {\bibinfo {volume}
  {7}},\ \bibinfo {pages} {100160} (\bibinfo {year} {2020})}\BibitemShut
  {NoStop}%
\bibitem [{\citenamefont {Ochoa}\ \emph {et~al.}(2020)\citenamefont {Ochoa},
  \citenamefont {Chicano},\ and\ \citenamefont {Tomassini}}]{Ochoa2020}%
  \BibitemOpen
  \bibfield  {author} {\bibinfo {author} {\bibfnamefont {G.}~\bibnamefont
  {Ochoa}}, \bibinfo {author} {\bibfnamefont {F.}~\bibnamefont {Chicano}}, \
  and\ \bibinfo {author} {\bibfnamefont {M.}~\bibnamefont {Tomassini}},\ }in\
  \href {\doibase 10.1007/978-3-030-58115-2_9} {\emph {\bibinfo {booktitle}
  {Parallel Problem Solving from Nature {\textendash} {PPSN} {XVI}}}}\
  (\bibinfo  {publisher} {Springer International Publishing},\ \bibinfo {year}
  {2020})\ pp.\ \bibinfo {pages} {125--138}\BibitemShut {NoStop}%
\bibitem [{\citenamefont {Garey}\ and\ \citenamefont
  {Johnson}(1990)}]{garey_johsnon_np}%
  \BibitemOpen
  \bibfield  {author} {\bibinfo {author} {\bibfnamefont {M.~R.}\ \bibnamefont
  {Garey}}\ and\ \bibinfo {author} {\bibfnamefont {D.~S.}\ \bibnamefont
  {Johnson}},\ }\href@noop {} {\emph {\bibinfo {title} {Computers and
  Intractability; A Guide to the Theory of NP-Completeness}}}\ (\bibinfo
  {publisher} {W. H. Freeman \& Co.},\ \bibinfo {address} {USA},\ \bibinfo
  {year} {1990})\BibitemShut {NoStop}%
\bibitem [{\citenamefont {Aspvall}\ \emph {et~al.}(1979)\citenamefont
  {Aspvall}, \citenamefont {Plass},\ and\ \citenamefont {Tarjan}}]{2sat-proof}%
  \BibitemOpen
  \bibfield  {author} {\bibinfo {author} {\bibfnamefont {B.}~\bibnamefont
  {Aspvall}}, \bibinfo {author} {\bibfnamefont {M.~F.}\ \bibnamefont {Plass}},
  \ and\ \bibinfo {author} {\bibfnamefont {R.~E.}\ \bibnamefont {Tarjan}},\
  }\href@noop {} {\bibfield  {journal} {\bibinfo  {journal} {Information
  Processing Letters}\ }\textbf {\bibinfo {volume} {8}},\ \bibinfo {pages}
  {121} (\bibinfo {year} {1979})}\BibitemShut {NoStop}%
\bibitem [{\citenamefont {Chvatal}\ and\ \citenamefont
  {Reed}(1992)}]{Chvatal1992}%
  \BibitemOpen
  \bibfield  {author} {\bibinfo {author} {\bibfnamefont {V.}~\bibnamefont
  {Chvatal}}\ and\ \bibinfo {author} {\bibfnamefont {B.}~\bibnamefont {Reed}},\
  }in\ \href {\doibase 10.1109/SFCS.1992.267789} {\emph {\bibinfo {booktitle}
  {Proceedings., 33rd Annual Symposium on Foundations of Computer Science}}}\
  (\bibinfo {year} {1992})\ pp.\ \bibinfo {pages} {620--627}\BibitemShut
  {NoStop}%
\bibitem [{\citenamefont {Papadimitriou}\ and\ \citenamefont
  {Steiglitz}(1998)}]{Papadimitriou1998-pc}%
  \BibitemOpen
  \bibfield  {author} {\bibinfo {author} {\bibfnamefont {C.~H.}\ \bibnamefont
  {Papadimitriou}}\ and\ \bibinfo {author} {\bibfnamefont {K.}~\bibnamefont
  {Steiglitz}},\ }\href@noop {} {\emph {\bibinfo {title} {Combinatorial
  optimization}}},\ Dover Books on Computer Science\ (\bibinfo  {publisher}
  {Dover Publications},\ \bibinfo {address} {Mineola, NY},\ \bibinfo {year}
  {1998})\BibitemShut {NoStop}%
\bibitem [{\citenamefont {Farhi}\ \emph {et~al.}(2000)\citenamefont {Farhi},
  \citenamefont {Goldstone}, \citenamefont {Gutmann},\ and\ \citenamefont
  {Sipser}}]{farhi_aqc}%
  \BibitemOpen
  \bibfield  {author} {\bibinfo {author} {\bibfnamefont {E.}~\bibnamefont
  {Farhi}}, \bibinfo {author} {\bibfnamefont {J.}~\bibnamefont {Goldstone}},
  \bibinfo {author} {\bibfnamefont {S.}~\bibnamefont {Gutmann}}, \ and\
  \bibinfo {author} {\bibfnamefont {M.}~\bibnamefont {Sipser}},\ }\href@noop {}
  {\enquote {\bibinfo {title} {Quantum computation by adiabatic evolution},}\ }
  (\bibinfo {year} {2000}),\ \Eprint {http://arxiv.org/abs/quant-ph/0001106}
  {arXiv:quant-ph/0001106} \BibitemShut {NoStop}%
\bibitem [{\citenamefont {Kadowaki}\ and\ \citenamefont
  {Nishimori}(1998)}]{Kadowaki_1998}%
  \BibitemOpen
  \bibfield  {author} {\bibinfo {author} {\bibfnamefont {T.}~\bibnamefont
  {Kadowaki}}\ and\ \bibinfo {author} {\bibfnamefont {H.}~\bibnamefont
  {Nishimori}},\ }\href {\doibase 10.1103/physreve.58.5355} {\bibfield
  {journal} {\bibinfo  {journal} {Physical Review E}\ }\textbf {\bibinfo
  {volume} {58}},\ \bibinfo {pages} {5355} (\bibinfo {year}
  {1998})}\BibitemShut {NoStop}%
\bibitem [{\citenamefont {Ozaeta}\ \emph {et~al.}(2022)\citenamefont {Ozaeta},
  \citenamefont {van Dam},\ and\ \citenamefont {McMahon}}]{Ozaeta2022-journal}%
  \BibitemOpen
  \bibfield  {author} {\bibinfo {author} {\bibfnamefont {A.}~\bibnamefont
  {Ozaeta}}, \bibinfo {author} {\bibfnamefont {W.}~\bibnamefont {van Dam}}, \
  and\ \bibinfo {author} {\bibfnamefont {P.~L.}\ \bibnamefont {McMahon}},\
  }\href {\doibase 10.1088/2058-9565/ac9013} {\bibfield  {journal} {\bibinfo
  {journal} {Quantum Science and Technology}\ }\textbf {\bibinfo {volume}
  {7}},\ \bibinfo {pages} {045036} (\bibinfo {year} {2022})}\BibitemShut
  {NoStop}%
\bibitem [{\citenamefont {Hauke}\ \emph {et~al.}(2020)\citenamefont {Hauke},
  \citenamefont {Katzgraber}, \citenamefont {Lechner}, \citenamefont
  {Nishimori},\ and\ \citenamefont {Oliver}}]{hauke_perspectives_2020}%
  \BibitemOpen
  \bibfield  {author} {\bibinfo {author} {\bibfnamefont {P.}~\bibnamefont
  {Hauke}}, \bibinfo {author} {\bibfnamefont {H.~G.}\ \bibnamefont
  {Katzgraber}}, \bibinfo {author} {\bibfnamefont {W.}~\bibnamefont {Lechner}},
  \bibinfo {author} {\bibfnamefont {H.}~\bibnamefont {Nishimori}}, \ and\
  \bibinfo {author} {\bibfnamefont {W.~D.}\ \bibnamefont {Oliver}},\ }\href
  {\doibase 10.1088/1361-6633/ab85b8} {\bibfield  {journal} {\bibinfo
  {journal} {Reports on Progress in Physics}\ }\textbf {\bibinfo {volume}
  {83}},\ \bibinfo {pages} {054401} (\bibinfo {year} {2020})}\BibitemShut
  {NoStop}%
\bibitem [{\citenamefont {Abram{\'{e}}}\ and\ \citenamefont
  {Habet}(2015)}]{Abrame2015}%
  \BibitemOpen
  \bibfield  {author} {\bibinfo {author} {\bibfnamefont {A.}~\bibnamefont
  {Abram{\'{e}}}}\ and\ \bibinfo {author} {\bibfnamefont {D.}~\bibnamefont
  {Habet}},\ }\href {\doibase 10.3233/sat190104} {\bibfield  {journal}
  {\bibinfo  {journal} {Journal on Satisfiability, Boolean Modeling and
  Computation}\ }\textbf {\bibinfo {volume} {9}},\ \bibinfo {pages} {89}
  (\bibinfo {year} {2015})}\BibitemShut {NoStop}%
\bibitem [{\citenamefont {Li}\ and\ \citenamefont {Many}(2021)}]{Li2021a}%
  \BibitemOpen
  \bibfield  {author} {\bibinfo {author} {\bibfnamefont {C.~M.}\ \bibnamefont
  {Li}}\ and\ \bibinfo {author} {\bibfnamefont {F.}~\bibnamefont {Many}},\
  }\href {\doibase 10.3233/978-1-58603-929-5-613} {\bibfield  {journal}
  {\bibinfo  {journal} {Frontiers in Artificial Intelligence and Applications}\
  }\textbf {\bibinfo {volume} {185}},\ \bibinfo {pages} {613} (\bibinfo {year}
  {2021})}\BibitemShut {NoStop}%
\bibitem [{\citenamefont {Borchers}\ and\ \citenamefont
  {Furman}(1998)}]{Borchers1998}%
  \BibitemOpen
  \bibfield  {author} {\bibinfo {author} {\bibfnamefont {B.}~\bibnamefont
  {Borchers}}\ and\ \bibinfo {author} {\bibfnamefont {J.}~\bibnamefont
  {Furman}},\ }\href {\doibase 10.1023/a:1009725216438} {\bibfield  {journal}
  {\bibinfo  {journal} {Journal of Combinatorial Optimization}\ }\textbf
  {\bibinfo {volume} {2}},\ \bibinfo {pages} {299} (\bibinfo {year}
  {1998})}\BibitemShut {NoStop}%
\bibitem [{\citenamefont {Patel}\ \emph {et~al.}(2022)\citenamefont {Patel},
  \citenamefont {Jerbi}, \citenamefont {Bäck},\ and\ \citenamefont
  {Dunjko}}]{patel_reinforcement_2022_1}%
  \BibitemOpen
  \bibfield  {author} {\bibinfo {author} {\bibfnamefont {Y.~J.}\ \bibnamefont
  {Patel}}, \bibinfo {author} {\bibfnamefont {S.}~\bibnamefont {Jerbi}},
  \bibinfo {author} {\bibfnamefont {T.}~\bibnamefont {Bäck}}, \ and\ \bibinfo
  {author} {\bibfnamefont {V.}~\bibnamefont {Dunjko}},\ }\href
  {http://arxiv.org/abs/2207.06294} {\  (\bibinfo {year} {2022})},\ \bibinfo
  {note} {arXiv:2207.06294 [quant-ph]}\BibitemShut {NoStop}%
\bibitem [{\citenamefont {Hukushima}\ and\ \citenamefont
  {Nemoto}(1996)}]{Hukushima_1996}%
  \BibitemOpen
  \bibfield  {author} {\bibinfo {author} {\bibfnamefont {K.}~\bibnamefont
  {Hukushima}}\ and\ \bibinfo {author} {\bibfnamefont {K.}~\bibnamefont
  {Nemoto}},\ }\href {\doibase 10.1143/jpsj.65.1604} {\bibfield  {journal}
  {\bibinfo  {journal} {Journal of the Physical Society of Japan}\ }\textbf
  {\bibinfo {volume} {65}},\ \bibinfo {pages} {1604} (\bibinfo {year}
  {1996})}\BibitemShut {NoStop}%
\bibitem [{\citenamefont {Earl}\ and\ \citenamefont {Deem}(2005)}]{Earl2005}%
  \BibitemOpen
  \bibfield  {author} {\bibinfo {author} {\bibfnamefont {D.~J.}\ \bibnamefont
  {Earl}}\ and\ \bibinfo {author} {\bibfnamefont {M.~W.}\ \bibnamefont
  {Deem}},\ }\href {\doibase 10.1039/b509983h} {\bibfield  {journal} {\bibinfo
  {journal} {Physical Chemistry Chemical Physics}\ }\textbf {\bibinfo {volume}
  {7}},\ \bibinfo {pages} {3910} (\bibinfo {year} {2005})}\BibitemShut
  {NoStop}%
\bibitem [{\citenamefont {Rom{\'{a}}}\ \emph {et~al.}(2009)\citenamefont
  {Rom{\'{a}}}, \citenamefont {Risau-Gusman}, \citenamefont {Ramirez-Pastor},
  \citenamefont {Nieto},\ and\ \citenamefont {Vogel}}]{Roma2009}%
  \BibitemOpen
  \bibfield  {author} {\bibinfo {author} {\bibfnamefont {F.}~\bibnamefont
  {Rom{\'{a}}}}, \bibinfo {author} {\bibfnamefont {S.}~\bibnamefont
  {Risau-Gusman}}, \bibinfo {author} {\bibfnamefont {A.}~\bibnamefont
  {Ramirez-Pastor}}, \bibinfo {author} {\bibfnamefont {F.}~\bibnamefont
  {Nieto}}, \ and\ \bibinfo {author} {\bibfnamefont {E.}~\bibnamefont
  {Vogel}},\ }\href {\doibase 10.1016/j.physa.2009.03.036} {\bibfield
  {journal} {\bibinfo  {journal} {Physica A: Statistical Mechanics and its
  Applications}\ }\textbf {\bibinfo {volume} {388}},\ \bibinfo {pages} {2821}
  (\bibinfo {year} {2009})}\BibitemShut {NoStop}%
\bibitem [{\citenamefont {Zhu}\ \emph {et~al.}(2020)\citenamefont {Zhu},
  \citenamefont {Fang},\ and\ \citenamefont
  {Katzgraber}}]{zhu_borealisgeneralized_2020}%
  \BibitemOpen
  \bibfield  {author} {\bibinfo {author} {\bibfnamefont {Z.}~\bibnamefont
  {Zhu}}, \bibinfo {author} {\bibfnamefont {C.}~\bibnamefont {Fang}}, \ and\
  \bibinfo {author} {\bibfnamefont {H.~G.}\ \bibnamefont {Katzgraber}},\ }\href
  {\doibase 10.1007/s11590-020-01570-7} {\bibfield  {journal} {\bibinfo
  {journal} {Optimization Letters}\ }\textbf {\bibinfo {volume} {14}},\
  \bibinfo {pages} {2495} (\bibinfo {year} {2020})}\BibitemShut {NoStop}%
\bibitem [{\citenamefont {Boettcher}\ and\ \citenamefont
  {Percus}(2003)}]{Boettcher_2003}%
  \BibitemOpen
  \bibfield  {author} {\bibinfo {author} {\bibfnamefont {S.}~\bibnamefont
  {Boettcher}}\ and\ \bibinfo {author} {\bibfnamefont {A.~G.}\ \bibnamefont
  {Percus}},\ }in\ \href {\doibase 10.1007/978-1-4615-1043-7_3} {\emph
  {\bibinfo {booktitle} {Computational Modeling and Problem Solving in the
  Networked World}}}\ (\bibinfo  {publisher} {Springer {US}},\ \bibinfo {year}
  {2003})\ pp.\ \bibinfo {pages} {61--77}\BibitemShut {NoStop}%
\bibitem [{\citenamefont {Mahajan}\ and\ \citenamefont
  {Slivovsky}(2023)}]{sat2023}%
  \BibitemOpen
  \bibfield  {author} {\bibinfo {author} {\bibfnamefont {M.}~\bibnamefont
  {Mahajan}}\ and\ \bibinfo {author} {\bibfnamefont {F.}~\bibnamefont
  {Slivovsky}},\ }in\ \href@noop {} {\emph {\bibinfo {booktitle} {26th
  International Conference on Theory and Applications of Satisfiability Testing
  (SAT 2023)}}}\ (\bibinfo {organization} {Schloss Dagstuhl-Leibniz-Zentrum
  f{\"u}r Informatik},\ \bibinfo {year} {2023})\BibitemShut {NoStop}%
\bibitem [{\citenamefont {Erd{\H o}s}\ and\ \citenamefont
  {R{\'e}nyi}(2022)}]{Erdos2022-hw}%
  \BibitemOpen
  \bibfield  {author} {\bibinfo {author} {\bibfnamefont {P.}~\bibnamefont
  {Erd{\H o}s}}\ and\ \bibinfo {author} {\bibfnamefont {A.}~\bibnamefont
  {R{\'e}nyi}},\ }\href@noop {} {\bibfield  {journal} {\bibinfo  {journal}
  {Publ. Math. Debrecen}\ }\textbf {\bibinfo {volume} {6}},\ \bibinfo {pages}
  {290} (\bibinfo {year} {2022})}\BibitemShut {NoStop}%
\bibitem [{\citenamefont {Dupont}\ \emph {et~al.}(2023)\citenamefont {Dupont},
  \citenamefont {Evert}, \citenamefont {Hodson}, \citenamefont {Sundar},
  \citenamefont {Jeffrey}, \citenamefont {Yamaguchi}, \citenamefont {Feng},
  \citenamefont {Maciejewski}, \citenamefont {Hadfield}, \citenamefont {Alam},
  \citenamefont {Wang}, \citenamefont {Grabbe}, \citenamefont {Lott},
  \citenamefont {Rieffel}, \citenamefont {Venturelli},\ and\ \citenamefont
  {Reagor}}]{dupont_quantum_2023}%
  \BibitemOpen
  \bibfield  {author} {\bibinfo {author} {\bibfnamefont {M.}~\bibnamefont
  {Dupont}}, \bibinfo {author} {\bibfnamefont {B.}~\bibnamefont {Evert}},
  \bibinfo {author} {\bibfnamefont {M.~J.}\ \bibnamefont {Hodson}}, \bibinfo
  {author} {\bibfnamefont {B.}~\bibnamefont {Sundar}}, \bibinfo {author}
  {\bibfnamefont {S.}~\bibnamefont {Jeffrey}}, \bibinfo {author} {\bibfnamefont
  {Y.}~\bibnamefont {Yamaguchi}}, \bibinfo {author} {\bibfnamefont
  {D.}~\bibnamefont {Feng}}, \bibinfo {author} {\bibfnamefont {F.~B.}\
  \bibnamefont {Maciejewski}}, \bibinfo {author} {\bibfnamefont
  {S.}~\bibnamefont {Hadfield}}, \bibinfo {author} {\bibfnamefont {M.~S.}\
  \bibnamefont {Alam}}, \bibinfo {author} {\bibfnamefont {Z.}~\bibnamefont
  {Wang}}, \bibinfo {author} {\bibfnamefont {S.}~\bibnamefont {Grabbe}},
  \bibinfo {author} {\bibfnamefont {P.~A.}\ \bibnamefont {Lott}}, \bibinfo
  {author} {\bibfnamefont {E.~G.}\ \bibnamefont {Rieffel}}, \bibinfo {author}
  {\bibfnamefont {D.}~\bibnamefont {Venturelli}}, \ and\ \bibinfo {author}
  {\bibfnamefont {M.~J.}\ \bibnamefont {Reagor}},\ }\href@noop {} {\bibfield
  {journal} {\bibinfo  {journal} {Science Advances}\ }\textbf {\bibinfo
  {volume} {9}},\ \bibinfo {pages} {eadi0487} (\bibinfo {year}
  {2023})}\BibitemShut {NoStop}%
\bibitem [{\citenamefont {Halldórsson}\ and\ \citenamefont
  {Radhakrishnan}(1997)}]{halldorsson_greed_1997}%
  \BibitemOpen
  \bibfield  {author} {\bibinfo {author} {\bibfnamefont {M.~M.}\ \bibnamefont
  {Halldórsson}}\ and\ \bibinfo {author} {\bibfnamefont {J.}~\bibnamefont
  {Radhakrishnan}},\ }\href {\doibase 10.1007/BF02523693} {\bibfield  {journal}
  {\bibinfo  {journal} {Algorithmica}\ }\textbf {\bibinfo {volume} {18}},\
  \bibinfo {pages} {145} (\bibinfo {year} {1997})}\BibitemShut {NoStop}%
\bibitem [{\citenamefont {B{\"{o}}ther}\ \emph {et~al.}(2022)\citenamefont
  {B{\"{o}}ther}, \citenamefont {Ki{\ss}ig}, \citenamefont {Taraz},
  \citenamefont {Cohen}, \citenamefont {Seidel},\ and\ \citenamefont
  {Friedrich}}]{boether}%
  \BibitemOpen
  \bibfield  {author} {\bibinfo {author} {\bibfnamefont {M.}~\bibnamefont
  {B{\"{o}}ther}}, \bibinfo {author} {\bibfnamefont {O.}~\bibnamefont
  {Ki{\ss}ig}}, \bibinfo {author} {\bibfnamefont {M.}~\bibnamefont {Taraz}},
  \bibinfo {author} {\bibfnamefont {S.}~\bibnamefont {Cohen}}, \bibinfo
  {author} {\bibfnamefont {K.}~\bibnamefont {Seidel}}, \ and\ \bibinfo {author}
  {\bibfnamefont {T.}~\bibnamefont {Friedrich}},\ }in\ \href
  {https://openreview.net/forum?id=mk0HzdqY7i1} {\emph {\bibinfo {booktitle}
  {The Tenth International Conference on Learning Representations, {ICLR} 2022,
  Virtual Event, April 25-29, 2022}}}\ (\bibinfo  {publisher}
  {OpenReview.net},\ \bibinfo {year} {2022})\BibitemShut {NoStop}%
\bibitem [{\citenamefont {Lamm}\ \emph {et~al.}(2017)\citenamefont {Lamm},
  \citenamefont {Sanders}, \citenamefont {Schulz}, \citenamefont {Strash},\
  and\ \citenamefont {Werneck}}]{lamm_finding_2017}%
  \BibitemOpen
  \bibfield  {author} {\bibinfo {author} {\bibfnamefont {S.}~\bibnamefont
  {Lamm}}, \bibinfo {author} {\bibfnamefont {P.}~\bibnamefont {Sanders}},
  \bibinfo {author} {\bibfnamefont {C.}~\bibnamefont {Schulz}}, \bibinfo
  {author} {\bibfnamefont {D.}~\bibnamefont {Strash}}, \ and\ \bibinfo {author}
  {\bibfnamefont {R.~F.}\ \bibnamefont {Werneck}},\ }\href {\doibase
  10.1007/s10732-017-9337-x} {\bibfield  {journal} {\bibinfo  {journal}
  {Journal of Heuristics}\ }\textbf {\bibinfo {volume} {23}},\ \bibinfo {pages}
  {207} (\bibinfo {year} {2017})}\BibitemShut {NoStop}%
\bibitem [{\citenamefont {Hespe}\ \emph {et~al.}(2019)\citenamefont {Hespe},
  \citenamefont {Schulz},\ and\ \citenamefont {Strash}}]{hespe_scalable_2019}%
  \BibitemOpen
  \bibfield  {author} {\bibinfo {author} {\bibfnamefont {D.}~\bibnamefont
  {Hespe}}, \bibinfo {author} {\bibfnamefont {C.}~\bibnamefont {Schulz}}, \
  and\ \bibinfo {author} {\bibfnamefont {D.}~\bibnamefont {Strash}},\ }\href
  {\doibase 10.1145/3355502} {\bibfield  {journal} {\bibinfo  {journal} {ACM
  Journal of Experimental Algorithmics}\ }\textbf {\bibinfo {volume} {24}},\
  \bibinfo {pages} {1} (\bibinfo {year} {2019})}\BibitemShut {NoStop}%
\bibitem [{\citenamefont {Lee}\ \emph {et~al.}(2021)\citenamefont {Lee},
  \citenamefont {Magann}, \citenamefont {Rabitz},\ and\ \citenamefont
  {Arenz}}]{lee_progress_2021}%
  \BibitemOpen
  \bibfield  {author} {\bibinfo {author} {\bibfnamefont {J.}~\bibnamefont
  {Lee}}, \bibinfo {author} {\bibfnamefont {A.~B.}\ \bibnamefont {Magann}},
  \bibinfo {author} {\bibfnamefont {H.~A.}\ \bibnamefont {Rabitz}}, \ and\
  \bibinfo {author} {\bibfnamefont {C.}~\bibnamefont {Arenz}},\ }\href
  {\doibase 10.1103/PhysRevA.104.032401} {\bibfield  {journal} {\bibinfo
  {journal} {Physical Review A}\ }\textbf {\bibinfo {volume} {104}},\ \bibinfo
  {pages} {032401} (\bibinfo {year} {2021})}\BibitemShut {NoStop}%
\bibitem [{\citenamefont {Bittel}\ and\ \citenamefont
  {Kliesch}(2021)}]{Bittel_2021}%
  \BibitemOpen
  \bibfield  {author} {\bibinfo {author} {\bibfnamefont {L.}~\bibnamefont
  {Bittel}}\ and\ \bibinfo {author} {\bibfnamefont {M.}~\bibnamefont
  {Kliesch}},\ }\href {\doibase 10.1103/physrevlett.127.120502} {\bibfield
  {journal} {\bibinfo  {journal} {Physical Review Letters}\ }\textbf {\bibinfo
  {volume} {127}} (\bibinfo {year} {2021}),\
  10.1103/physrevlett.127.120502}\BibitemShut {NoStop}%
\bibitem [{\citenamefont {Liu}\ \emph {et~al.}(2023)\citenamefont {Liu},
  \citenamefont {Gao}, \citenamefont {Cain}, \citenamefont {Lukin},\ and\
  \citenamefont {Wang}}]{liu_computing_2023}%
  \BibitemOpen
  \bibfield  {author} {\bibinfo {author} {\bibfnamefont {J.-G.}\ \bibnamefont
  {Liu}}, \bibinfo {author} {\bibfnamefont {X.}~\bibnamefont {Gao}}, \bibinfo
  {author} {\bibfnamefont {M.}~\bibnamefont {Cain}}, \bibinfo {author}
  {\bibfnamefont {M.~D.}\ \bibnamefont {Lukin}}, \ and\ \bibinfo {author}
  {\bibfnamefont {S.-T.}\ \bibnamefont {Wang}},\ }\href {\doibase
  10.1137/22M1501787} {\bibfield  {journal} {\bibinfo  {journal} {SIAM Journal
  on Scientific Computing}\ }\textbf {\bibinfo {volume} {45}},\ \bibinfo
  {pages} {A1239} (\bibinfo {year} {2023})}\BibitemShut {NoStop}%
\bibitem [{\citenamefont {Finžgar}\ \emph {et~al.}(2023)\citenamefont
  {Finžgar}, \citenamefont {Schuetz}, \citenamefont {Brubaker}, \citenamefont
  {Nishimori},\ and\ \citenamefont {Katzgraber}}]{finzgar_bo_2023}%
  \BibitemOpen
  \bibfield  {author} {\bibinfo {author} {\bibfnamefont {J.~R.}\ \bibnamefont
  {Finžgar}}, \bibinfo {author} {\bibfnamefont {M.~J.~A.}\ \bibnamefont
  {Schuetz}}, \bibinfo {author} {\bibfnamefont {J.~K.}\ \bibnamefont
  {Brubaker}}, \bibinfo {author} {\bibfnamefont {H.}~\bibnamefont {Nishimori}},
  \ and\ \bibinfo {author} {\bibfnamefont {H.~G.}\ \bibnamefont {Katzgraber}},\
  }\href@noop {} {\  (\bibinfo {year} {2023})},\ \Eprint
  {http://arxiv.org/abs/2305.13365} {arXiv:2305.13365} \BibitemShut {NoStop}%
\bibitem [{\citenamefont {Kirkpatrick}\ \emph {et~al.}(1983)\citenamefont
  {Kirkpatrick}, \citenamefont {Gelatt},\ and\ \citenamefont
  {Vecchi}}]{kirkpatrick-sa}%
  \BibitemOpen
  \bibfield  {author} {\bibinfo {author} {\bibfnamefont {S.}~\bibnamefont
  {Kirkpatrick}}, \bibinfo {author} {\bibfnamefont {C.~D.}\ \bibnamefont
  {Gelatt}}, \ and\ \bibinfo {author} {\bibfnamefont {M.~P.}\ \bibnamefont
  {Vecchi}},\ }\href {\doibase 10.1126/science.220.4598.671} {\bibfield
  {journal} {\bibinfo  {journal} {Science}\ }\textbf {\bibinfo {volume}
  {220}},\ \bibinfo {pages} {671} (\bibinfo {year} {1983})}\BibitemShut
  {NoStop}%
\bibitem [{\citenamefont {Ignatiev}\ \emph {et~al.}(2019)\citenamefont
  {Ignatiev}, \citenamefont {Morgado},\ and\ \citenamefont
  {Marques-Silva}}]{Ignatiev2019}%
  \BibitemOpen
  \bibfield  {author} {\bibinfo {author} {\bibfnamefont {A.}~\bibnamefont
  {Ignatiev}}, \bibinfo {author} {\bibfnamefont {A.}~\bibnamefont {Morgado}}, \
  and\ \bibinfo {author} {\bibfnamefont {J.}~\bibnamefont {Marques-Silva}},\
  }\href {\doibase 10.3233/sat190116} {\bibfield  {journal} {\bibinfo
  {journal} {Journal on Satisfiability, Boolean Modeling and Computation}\
  }\textbf {\bibinfo {volume} {11}},\ \bibinfo {pages} {53} (\bibinfo {year}
  {2019})}\BibitemShut {NoStop}%
\bibitem [{\citenamefont {Hadfield}\ \emph {et~al.}(2019)\citenamefont
  {Hadfield}, \citenamefont {Wang}, \citenamefont {O'Gorman}, \citenamefont
  {Rieffel}, \citenamefont {Venturelli},\ and\ \citenamefont
  {Biswas}}]{hadfield_quantum_2019}%
  \BibitemOpen
  \bibfield  {author} {\bibinfo {author} {\bibfnamefont {S.}~\bibnamefont
  {Hadfield}}, \bibinfo {author} {\bibfnamefont {Z.}~\bibnamefont {Wang}},
  \bibinfo {author} {\bibfnamefont {B.}~\bibnamefont {O'Gorman}}, \bibinfo
  {author} {\bibfnamefont {E.}~\bibnamefont {Rieffel}}, \bibinfo {author}
  {\bibfnamefont {D.}~\bibnamefont {Venturelli}}, \ and\ \bibinfo {author}
  {\bibfnamefont {R.}~\bibnamefont {Biswas}},\ }\href {\doibase
  10.3390/a12020034} {\bibfield  {journal} {\bibinfo  {journal} {Algorithms}\
  }\textbf {\bibinfo {volume} {12}},\ \bibinfo {pages} {34} (\bibinfo {year}
  {2019})}\BibitemShut {NoStop}%
\bibitem [{\citenamefont {Hastings}(2019)}]{Hastings2019}%
  \BibitemOpen
  \bibfield  {author} {\bibinfo {author} {\bibfnamefont {M.~B.}\ \bibnamefont
  {Hastings}},\ }\href@noop {} {\  (\bibinfo {year} {2019})},\ \Eprint
  {http://arxiv.org/abs/1905.07047} {arXiv:1905.07047} \BibitemShut {NoStop}%
\bibitem [{\citenamefont {Huang}\ \emph {et~al.}(2021)\citenamefont {Huang},
  \citenamefont {Zhang}, \citenamefont {Newman}, \citenamefont {Ni},
  \citenamefont {Ding}, \citenamefont {Cai}, \citenamefont {Gao}, \citenamefont
  {Wang}, \citenamefont {Wu}, \citenamefont {Zhang}, \citenamefont {Ku},
  \citenamefont {Tian}, \citenamefont {Wu}, \citenamefont {Xu}, \citenamefont
  {Yu}, \citenamefont {Yuan}, \citenamefont {Szegedy}, \citenamefont {Shi},
  \citenamefont {Zhao}, \citenamefont {Deng},\ and\ \citenamefont
  {Chen}}]{Huang2021}%
  \BibitemOpen
  \bibfield  {author} {\bibinfo {author} {\bibfnamefont {C.}~\bibnamefont
  {Huang}}, \bibinfo {author} {\bibfnamefont {F.}~\bibnamefont {Zhang}},
  \bibinfo {author} {\bibfnamefont {M.}~\bibnamefont {Newman}}, \bibinfo
  {author} {\bibfnamefont {X.}~\bibnamefont {Ni}}, \bibinfo {author}
  {\bibfnamefont {D.}~\bibnamefont {Ding}}, \bibinfo {author} {\bibfnamefont
  {J.}~\bibnamefont {Cai}}, \bibinfo {author} {\bibfnamefont {X.}~\bibnamefont
  {Gao}}, \bibinfo {author} {\bibfnamefont {T.}~\bibnamefont {Wang}}, \bibinfo
  {author} {\bibfnamefont {F.}~\bibnamefont {Wu}}, \bibinfo {author}
  {\bibfnamefont {G.}~\bibnamefont {Zhang}}, \bibinfo {author} {\bibfnamefont
  {H.-S.}\ \bibnamefont {Ku}}, \bibinfo {author} {\bibfnamefont
  {Z.}~\bibnamefont {Tian}}, \bibinfo {author} {\bibfnamefont {J.}~\bibnamefont
  {Wu}}, \bibinfo {author} {\bibfnamefont {H.}~\bibnamefont {Xu}}, \bibinfo
  {author} {\bibfnamefont {H.}~\bibnamefont {Yu}}, \bibinfo {author}
  {\bibfnamefont {B.}~\bibnamefont {Yuan}}, \bibinfo {author} {\bibfnamefont
  {M.}~\bibnamefont {Szegedy}}, \bibinfo {author} {\bibfnamefont
  {Y.}~\bibnamefont {Shi}}, \bibinfo {author} {\bibfnamefont {H.-H.}\
  \bibnamefont {Zhao}}, \bibinfo {author} {\bibfnamefont {C.}~\bibnamefont
  {Deng}}, \ and\ \bibinfo {author} {\bibfnamefont {J.}~\bibnamefont {Chen}},\
  }\href {\doibase 10.1038/s43588-021-00119-7} {\bibfield  {journal} {\bibinfo
  {journal} {Nature Computational Science}\ }\textbf {\bibinfo {volume} {1}},\
  \bibinfo {pages} {578} (\bibinfo {year} {2021})}\BibitemShut {NoStop}%
\bibitem [{\citenamefont {Gray}\ and\ \citenamefont
  {Kourtis}(2021)}]{Gray_2021}%
  \BibitemOpen
  \bibfield  {author} {\bibinfo {author} {\bibfnamefont {J.}~\bibnamefont
  {Gray}}\ and\ \bibinfo {author} {\bibfnamefont {S.}~\bibnamefont {Kourtis}},\
  }\href {\doibase 10.22331/q-2021-03-15-410} {\bibfield  {journal} {\bibinfo
  {journal} {Quantum}\ }\textbf {\bibinfo {volume} {5}},\ \bibinfo {pages}
  {410} (\bibinfo {year} {2021})}\BibitemShut {NoStop}%
\bibitem [{\citenamefont {Johnson}\ \emph {et~al.}(2011)\citenamefont
  {Johnson}, \citenamefont {Amin}, \citenamefont {Gildert}, \citenamefont
  {Lanting}, \citenamefont {Hamze}, \citenamefont {Dickson}, \citenamefont
  {Harris}, \citenamefont {Berkley}, \citenamefont {Johansson}, \citenamefont
  {Bunyk}, \citenamefont {Chapple}, \citenamefont {Enderud}, \citenamefont
  {Hilton}, \citenamefont {Karimi}, \citenamefont {Ladizinsky}, \citenamefont
  {Ladizinsky}, \citenamefont {Oh}, \citenamefont {Perminov}, \citenamefont
  {Rich}, \citenamefont {Thom}, \citenamefont {Tolkacheva}, \citenamefont
  {Truncik}, \citenamefont {Uchaikin}, \citenamefont {Wang}, \citenamefont
  {Wilson},\ and\ \citenamefont {Rose}}]{Johnson_2011}%
  \BibitemOpen
  \bibfield  {author} {\bibinfo {author} {\bibfnamefont {M.~W.}\ \bibnamefont
  {Johnson}}, \bibinfo {author} {\bibfnamefont {M.~H.~S.}\ \bibnamefont
  {Amin}}, \bibinfo {author} {\bibfnamefont {S.}~\bibnamefont {Gildert}},
  \bibinfo {author} {\bibfnamefont {T.}~\bibnamefont {Lanting}}, \bibinfo
  {author} {\bibfnamefont {F.}~\bibnamefont {Hamze}}, \bibinfo {author}
  {\bibfnamefont {N.}~\bibnamefont {Dickson}}, \bibinfo {author} {\bibfnamefont
  {R.}~\bibnamefont {Harris}}, \bibinfo {author} {\bibfnamefont {A.~J.}\
  \bibnamefont {Berkley}}, \bibinfo {author} {\bibfnamefont {J.}~\bibnamefont
  {Johansson}}, \bibinfo {author} {\bibfnamefont {P.}~\bibnamefont {Bunyk}},
  \bibinfo {author} {\bibfnamefont {E.~M.}\ \bibnamefont {Chapple}}, \bibinfo
  {author} {\bibfnamefont {C.}~\bibnamefont {Enderud}}, \bibinfo {author}
  {\bibfnamefont {J.~P.}\ \bibnamefont {Hilton}}, \bibinfo {author}
  {\bibfnamefont {K.}~\bibnamefont {Karimi}}, \bibinfo {author} {\bibfnamefont
  {E.}~\bibnamefont {Ladizinsky}}, \bibinfo {author} {\bibfnamefont
  {N.}~\bibnamefont {Ladizinsky}}, \bibinfo {author} {\bibfnamefont
  {T.}~\bibnamefont {Oh}}, \bibinfo {author} {\bibfnamefont {I.}~\bibnamefont
  {Perminov}}, \bibinfo {author} {\bibfnamefont {C.}~\bibnamefont {Rich}},
  \bibinfo {author} {\bibfnamefont {M.~C.}\ \bibnamefont {Thom}}, \bibinfo
  {author} {\bibfnamefont {E.}~\bibnamefont {Tolkacheva}}, \bibinfo {author}
  {\bibfnamefont {C.~J.~S.}\ \bibnamefont {Truncik}}, \bibinfo {author}
  {\bibfnamefont {S.}~\bibnamefont {Uchaikin}}, \bibinfo {author}
  {\bibfnamefont {J.}~\bibnamefont {Wang}}, \bibinfo {author} {\bibfnamefont
  {B.}~\bibnamefont {Wilson}}, \ and\ \bibinfo {author} {\bibfnamefont
  {G.}~\bibnamefont {Rose}},\ }\href {\doibase 10.1038/nature10012} {\bibfield
  {journal} {\bibinfo  {journal} {Nature}\ }\textbf {\bibinfo {volume} {473}},\
  \bibinfo {pages} {194} (\bibinfo {year} {2011})}\BibitemShut {NoStop}%
\bibitem [{\citenamefont {King}\ \emph {et~al.}(2022)\citenamefont {King},
  \citenamefont {Suzuki}, \citenamefont {Raymond}, \citenamefont {Zucca},
  \citenamefont {Lanting}, \citenamefont {Altomare}, \citenamefont {Berkley},
  \citenamefont {Ejtemaee}, \citenamefont {Hoskinson}, \citenamefont {Huang},
  \citenamefont {Ladizinsky}, \citenamefont {MacDonald}, \citenamefont
  {Marsden}, \citenamefont {Oh}, \citenamefont {Poulin-Lamarre}, \citenamefont
  {Reis}, \citenamefont {Rich}, \citenamefont {Sato}, \citenamefont
  {Whittaker}, \citenamefont {Yao}, \citenamefont {Harris}, \citenamefont
  {Lidar}, \citenamefont {Nishimori},\ and\ \citenamefont {Amin}}]{King_2022}%
  \BibitemOpen
  \bibfield  {author} {\bibinfo {author} {\bibfnamefont {A.~D.}\ \bibnamefont
  {King}}, \bibinfo {author} {\bibfnamefont {S.}~\bibnamefont {Suzuki}},
  \bibinfo {author} {\bibfnamefont {J.}~\bibnamefont {Raymond}}, \bibinfo
  {author} {\bibfnamefont {A.}~\bibnamefont {Zucca}}, \bibinfo {author}
  {\bibfnamefont {T.}~\bibnamefont {Lanting}}, \bibinfo {author} {\bibfnamefont
  {F.}~\bibnamefont {Altomare}}, \bibinfo {author} {\bibfnamefont {A.~J.}\
  \bibnamefont {Berkley}}, \bibinfo {author} {\bibfnamefont {S.}~\bibnamefont
  {Ejtemaee}}, \bibinfo {author} {\bibfnamefont {E.}~\bibnamefont {Hoskinson}},
  \bibinfo {author} {\bibfnamefont {S.}~\bibnamefont {Huang}}, \bibinfo
  {author} {\bibfnamefont {E.}~\bibnamefont {Ladizinsky}}, \bibinfo {author}
  {\bibfnamefont {A.~J.~R.}\ \bibnamefont {MacDonald}}, \bibinfo {author}
  {\bibfnamefont {G.}~\bibnamefont {Marsden}}, \bibinfo {author} {\bibfnamefont
  {T.}~\bibnamefont {Oh}}, \bibinfo {author} {\bibfnamefont {G.}~\bibnamefont
  {Poulin-Lamarre}}, \bibinfo {author} {\bibfnamefont {M.}~\bibnamefont
  {Reis}}, \bibinfo {author} {\bibfnamefont {C.}~\bibnamefont {Rich}}, \bibinfo
  {author} {\bibfnamefont {Y.}~\bibnamefont {Sato}}, \bibinfo {author}
  {\bibfnamefont {J.~D.}\ \bibnamefont {Whittaker}}, \bibinfo {author}
  {\bibfnamefont {J.}~\bibnamefont {Yao}}, \bibinfo {author} {\bibfnamefont
  {R.}~\bibnamefont {Harris}}, \bibinfo {author} {\bibfnamefont {D.~A.}\
  \bibnamefont {Lidar}}, \bibinfo {author} {\bibfnamefont {H.}~\bibnamefont
  {Nishimori}}, \ and\ \bibinfo {author} {\bibfnamefont {M.~H.}\ \bibnamefont
  {Amin}},\ }\href {\doibase 10.1038/s41567-022-01741-6} {\bibfield  {journal}
  {\bibinfo  {journal} {Nature Physics}\ }\textbf {\bibinfo {volume} {18}},\
  \bibinfo {pages} {1324} (\bibinfo {year} {2022})}\BibitemShut {NoStop}%
\bibitem [{\citenamefont {Caha}\ \emph {et~al.}(2022)\citenamefont {Caha},
  \citenamefont {Kliesch},\ and\ \citenamefont {Koenig}}]{caha_twisted_2022}%
  \BibitemOpen
  \bibfield  {author} {\bibinfo {author} {\bibfnamefont {L.}~\bibnamefont
  {Caha}}, \bibinfo {author} {\bibfnamefont {A.}~\bibnamefont {Kliesch}}, \
  and\ \bibinfo {author} {\bibfnamefont {R.}~\bibnamefont {Koenig}},\ }\href
  {\doibase 10.1088/2058-9565/ac7f4f} {\bibfield  {journal} {\bibinfo
  {journal} {Quantum Science and Technology}\ }\textbf {\bibinfo {volume}
  {7}},\ \bibinfo {pages} {045013} (\bibinfo {year} {2022})}\BibitemShut
  {NoStop}%
\bibitem [{\citenamefont {Wagner}\ \emph {et~al.}(2023)\citenamefont {Wagner},
  \citenamefont {Nüßlein},\ and\ \citenamefont
  {Liers}}]{wagner2023enhancing}%
  \BibitemOpen
  \bibfield  {author} {\bibinfo {author} {\bibfnamefont {F.}~\bibnamefont
  {Wagner}}, \bibinfo {author} {\bibfnamefont {J.}~\bibnamefont {Nüßlein}}, \
  and\ \bibinfo {author} {\bibfnamefont {F.}~\bibnamefont {Liers}},\
  }\href@noop {} {\  (\bibinfo {year} {2023})},\ \Eprint
  {http://arxiv.org/abs/2302.05493} {arXiv:2302.05493} \BibitemShut {NoStop}%
\bibitem [{Note1()}]{Note1}%
  \BibitemOpen
  \bibinfo {note} {\protect \href
  {https://github.com/jernejrudifinzgar/qiro}{https://github.com/jernejrudifinzgar/qiro}}\BibitemShut
  {NoStop}%
\bibitem [{\citenamefont {Farhi}\ and\ \citenamefont
  {Harrow}(2019)}]{Farhi2016}%
  \BibitemOpen
  \bibfield  {author} {\bibinfo {author} {\bibfnamefont {E.}~\bibnamefont
  {Farhi}}\ and\ \bibinfo {author} {\bibfnamefont {A.~W.}\ \bibnamefont
  {Harrow}},\ }\href@noop {} {\  (\bibinfo {year} {2019})},\ \Eprint
  {http://arxiv.org/abs/1602.07674} {arXiv:1602.07674} \BibitemShut {NoStop}%
\bibitem [{\citenamefont {Bae}\ and\ \citenamefont
  {Lee}(2023)}]{bae_recursive_2022}%
  \BibitemOpen
  \bibfield  {author} {\bibinfo {author} {\bibfnamefont {E.}~\bibnamefont
  {Bae}}\ and\ \bibinfo {author} {\bibfnamefont {S.}~\bibnamefont {Lee}},\
  }\href@noop {} {\  (\bibinfo {year} {2023})},\ \Eprint
  {http://arxiv.org/abs/2211.15832} {arXiv:2211.15832} \BibitemShut {NoStop}%
\bibitem [{\citenamefont {Niedermeier}\ and\ \citenamefont
  {Rossmanith}(2000)}]{Niedermeier2000}%
  \BibitemOpen
  \bibfield  {author} {\bibinfo {author} {\bibfnamefont {R.}~\bibnamefont
  {Niedermeier}}\ and\ \bibinfo {author} {\bibfnamefont {P.}~\bibnamefont
  {Rossmanith}},\ }\href {\doibase 10.1006/jagm.2000.1075} {\bibfield
  {journal} {\bibinfo  {journal} {Journal of Algorithms}\ }\textbf {\bibinfo
  {volume} {36}},\ \bibinfo {pages} {63} (\bibinfo {year} {2000})}\BibitemShut
  {NoStop}%
\bibitem [{\citenamefont {Andrist}\ \emph {et~al.}(2023)\citenamefont
  {Andrist}, \citenamefont {Schuetz}, \citenamefont {Minssen}, \citenamefont
  {Yalovetzky}, \citenamefont {Chakrabarti}, \citenamefont {Herman},
  \citenamefont {Kumar}, \citenamefont {Salton}, \citenamefont {Shaydulin},
  \citenamefont {Sun}, \citenamefont {Pistoia},\ and\ \citenamefont
  {Katzgraber}}]{andrist_hardness_2023_1}%
  \BibitemOpen
  \bibfield  {author} {\bibinfo {author} {\bibfnamefont {R.~S.}\ \bibnamefont
  {Andrist}}, \bibinfo {author} {\bibfnamefont {M.~J.~A.}\ \bibnamefont
  {Schuetz}}, \bibinfo {author} {\bibfnamefont {P.}~\bibnamefont {Minssen}},
  \bibinfo {author} {\bibfnamefont {R.}~\bibnamefont {Yalovetzky}}, \bibinfo
  {author} {\bibfnamefont {S.}~\bibnamefont {Chakrabarti}}, \bibinfo {author}
  {\bibfnamefont {D.}~\bibnamefont {Herman}}, \bibinfo {author} {\bibfnamefont
  {N.}~\bibnamefont {Kumar}}, \bibinfo {author} {\bibfnamefont
  {G.}~\bibnamefont {Salton}}, \bibinfo {author} {\bibfnamefont
  {R.}~\bibnamefont {Shaydulin}}, \bibinfo {author} {\bibfnamefont
  {Y.}~\bibnamefont {Sun}}, \bibinfo {author} {\bibfnamefont {M.}~\bibnamefont
  {Pistoia}}, \ and\ \bibinfo {author} {\bibfnamefont {H.~G.}\ \bibnamefont
  {Katzgraber}},\ }\href@noop {} {\  (\bibinfo {year} {2023})},\ \Eprint
  {http://arxiv.org/abs/2307.09442} {arXiv:2307.09442} \BibitemShut {NoStop}%
\bibitem [{\citenamefont {Liu}\ \emph {et~al.}(2021)\citenamefont {Liu},
  \citenamefont {Wang},\ and\ \citenamefont {Zhang}}]{liu_tropical_2021}%
  \BibitemOpen
  \bibfield  {author} {\bibinfo {author} {\bibfnamefont {J.-G.}\ \bibnamefont
  {Liu}}, \bibinfo {author} {\bibfnamefont {L.}~\bibnamefont {Wang}}, \ and\
  \bibinfo {author} {\bibfnamefont {P.}~\bibnamefont {Zhang}},\ }\href
  {\doibase 10.1103/PhysRevLett.126.090506} {\bibfield  {journal} {\bibinfo
  {journal} {Physical Review Letters}\ }\textbf {\bibinfo {volume} {126}},\
  \bibinfo {pages} {090506} (\bibinfo {year} {2021})}\BibitemShut {NoStop}%
\bibitem [{\citenamefont {Bergholm}\ \emph {et~al.}(2022)\citenamefont
  {Bergholm}, \citenamefont {Izaac}, \citenamefont {Schuld}, \citenamefont
  {Gogolin}, \citenamefont {Ahmed}, \citenamefont {Ajith}, \citenamefont
  {Alam}, \citenamefont {Alonso-Linaje}, \citenamefont {AkashNarayanan},
  \citenamefont {Asadi}, \citenamefont {Arrazola}, \citenamefont {Azad},
  \citenamefont {Banning}, \citenamefont {Blank}, \citenamefont {Bromley},
  \citenamefont {Cordier}, \citenamefont {Ceroni}, \citenamefont {Delgado},
  \citenamefont {Matteo}, \citenamefont {Dusko}, \citenamefont {Garg},
  \citenamefont {Guala}, \citenamefont {Hayes}, \citenamefont {Hill},
  \citenamefont {Ijaz}, \citenamefont {Isacsson}, \citenamefont {Ittah},
  \citenamefont {Jahangiri}, \citenamefont {Jain}, \citenamefont {Jiang},
  \citenamefont {Khandelwal}, \citenamefont {Kottmann}, \citenamefont {Lang},
  \citenamefont {Lee}, \citenamefont {Loke}, \citenamefont {Lowe},
  \citenamefont {McKiernan}, \citenamefont {Meyer}, \citenamefont
  {Montañez-Barrera}, \citenamefont {Moyard}, \citenamefont {Niu},
  \citenamefont {O'Riordan}, \citenamefont {Oud}, \citenamefont {Panigrahi},
  \citenamefont {Park}, \citenamefont {Polatajko}, \citenamefont {Quesada},
  \citenamefont {Roberts}, \citenamefont {Sá}, \citenamefont {Schoch},
  \citenamefont {Shi}, \citenamefont {Shu}, \citenamefont {Sim}, \citenamefont
  {Singh}, \citenamefont {Strandberg}, \citenamefont {Soni}, \citenamefont
  {Száva}, \citenamefont {Thabet}, \citenamefont {Vargas-Hernández},
  \citenamefont {Vincent}, \citenamefont {Vitucci}, \citenamefont {Weber},
  \citenamefont {Wierichs}, \citenamefont {Wiersema}, \citenamefont {Willmann},
  \citenamefont {Wong}, \citenamefont {Zhang},\ and\ \citenamefont
  {Killoran}}]{bergholm2022pennylane}%
  \BibitemOpen
  \bibfield  {author} {\bibinfo {author} {\bibfnamefont {V.}~\bibnamefont
  {Bergholm}}, \bibinfo {author} {\bibfnamefont {J.}~\bibnamefont {Izaac}},
  \bibinfo {author} {\bibfnamefont {M.}~\bibnamefont {Schuld}}, \bibinfo
  {author} {\bibfnamefont {C.}~\bibnamefont {Gogolin}}, \bibinfo {author}
  {\bibfnamefont {S.}~\bibnamefont {Ahmed}}, \bibinfo {author} {\bibfnamefont
  {V.}~\bibnamefont {Ajith}}, \bibinfo {author} {\bibfnamefont {M.~S.}\
  \bibnamefont {Alam}}, \bibinfo {author} {\bibfnamefont {G.}~\bibnamefont
  {Alonso-Linaje}}, \bibinfo {author} {\bibfnamefont {B.}~\bibnamefont
  {AkashNarayanan}}, \bibinfo {author} {\bibfnamefont {A.}~\bibnamefont
  {Asadi}}, \bibinfo {author} {\bibfnamefont {J.~M.}\ \bibnamefont {Arrazola}},
  \bibinfo {author} {\bibfnamefont {U.}~\bibnamefont {Azad}}, \bibinfo {author}
  {\bibfnamefont {S.}~\bibnamefont {Banning}}, \bibinfo {author} {\bibfnamefont
  {C.}~\bibnamefont {Blank}}, \bibinfo {author} {\bibfnamefont {T.~R.}\
  \bibnamefont {Bromley}}, \bibinfo {author} {\bibfnamefont {B.~A.}\
  \bibnamefont {Cordier}}, \bibinfo {author} {\bibfnamefont {J.}~\bibnamefont
  {Ceroni}}, \bibinfo {author} {\bibfnamefont {A.}~\bibnamefont {Delgado}},
  \bibinfo {author} {\bibfnamefont {O.~D.}\ \bibnamefont {Matteo}}, \bibinfo
  {author} {\bibfnamefont {A.}~\bibnamefont {Dusko}}, \bibinfo {author}
  {\bibfnamefont {T.}~\bibnamefont {Garg}}, \bibinfo {author} {\bibfnamefont
  {D.}~\bibnamefont {Guala}}, \bibinfo {author} {\bibfnamefont
  {A.}~\bibnamefont {Hayes}}, \bibinfo {author} {\bibfnamefont
  {R.}~\bibnamefont {Hill}}, \bibinfo {author} {\bibfnamefont {A.}~\bibnamefont
  {Ijaz}}, \bibinfo {author} {\bibfnamefont {T.}~\bibnamefont {Isacsson}},
  \bibinfo {author} {\bibfnamefont {D.}~\bibnamefont {Ittah}}, \bibinfo
  {author} {\bibfnamefont {S.}~\bibnamefont {Jahangiri}}, \bibinfo {author}
  {\bibfnamefont {P.}~\bibnamefont {Jain}}, \bibinfo {author} {\bibfnamefont
  {E.}~\bibnamefont {Jiang}}, \bibinfo {author} {\bibfnamefont
  {A.}~\bibnamefont {Khandelwal}}, \bibinfo {author} {\bibfnamefont
  {K.}~\bibnamefont {Kottmann}}, \bibinfo {author} {\bibfnamefont {R.~A.}\
  \bibnamefont {Lang}}, \bibinfo {author} {\bibfnamefont {C.}~\bibnamefont
  {Lee}}, \bibinfo {author} {\bibfnamefont {T.}~\bibnamefont {Loke}}, \bibinfo
  {author} {\bibfnamefont {A.}~\bibnamefont {Lowe}}, \bibinfo {author}
  {\bibfnamefont {K.}~\bibnamefont {McKiernan}}, \bibinfo {author}
  {\bibfnamefont {J.~J.}\ \bibnamefont {Meyer}}, \bibinfo {author}
  {\bibfnamefont {J.~A.}\ \bibnamefont {Montañez-Barrera}}, \bibinfo {author}
  {\bibfnamefont {R.}~\bibnamefont {Moyard}}, \bibinfo {author} {\bibfnamefont
  {Z.}~\bibnamefont {Niu}}, \bibinfo {author} {\bibfnamefont {L.~J.}\
  \bibnamefont {O'Riordan}}, \bibinfo {author} {\bibfnamefont {S.}~\bibnamefont
  {Oud}}, \bibinfo {author} {\bibfnamefont {A.}~\bibnamefont {Panigrahi}},
  \bibinfo {author} {\bibfnamefont {C.-Y.}\ \bibnamefont {Park}}, \bibinfo
  {author} {\bibfnamefont {D.}~\bibnamefont {Polatajko}}, \bibinfo {author}
  {\bibfnamefont {N.}~\bibnamefont {Quesada}}, \bibinfo {author} {\bibfnamefont
  {C.}~\bibnamefont {Roberts}}, \bibinfo {author} {\bibfnamefont
  {N.}~\bibnamefont {Sá}}, \bibinfo {author} {\bibfnamefont {I.}~\bibnamefont
  {Schoch}}, \bibinfo {author} {\bibfnamefont {B.}~\bibnamefont {Shi}},
  \bibinfo {author} {\bibfnamefont {S.}~\bibnamefont {Shu}}, \bibinfo {author}
  {\bibfnamefont {S.}~\bibnamefont {Sim}}, \bibinfo {author} {\bibfnamefont
  {A.}~\bibnamefont {Singh}}, \bibinfo {author} {\bibfnamefont
  {I.}~\bibnamefont {Strandberg}}, \bibinfo {author} {\bibfnamefont
  {J.}~\bibnamefont {Soni}}, \bibinfo {author} {\bibfnamefont {A.}~\bibnamefont
  {Száva}}, \bibinfo {author} {\bibfnamefont {S.}~\bibnamefont {Thabet}},
  \bibinfo {author} {\bibfnamefont {R.~A.}\ \bibnamefont {Vargas-Hernández}},
  \bibinfo {author} {\bibfnamefont {T.}~\bibnamefont {Vincent}}, \bibinfo
  {author} {\bibfnamefont {N.}~\bibnamefont {Vitucci}}, \bibinfo {author}
  {\bibfnamefont {M.}~\bibnamefont {Weber}}, \bibinfo {author} {\bibfnamefont
  {D.}~\bibnamefont {Wierichs}}, \bibinfo {author} {\bibfnamefont
  {R.}~\bibnamefont {Wiersema}}, \bibinfo {author} {\bibfnamefont
  {M.}~\bibnamefont {Willmann}}, \bibinfo {author} {\bibfnamefont
  {V.}~\bibnamefont {Wong}}, \bibinfo {author} {\bibfnamefont {S.}~\bibnamefont
  {Zhang}}, \ and\ \bibinfo {author} {\bibfnamefont {N.}~\bibnamefont
  {Killoran}},\ }\href@noop {} {\  (\bibinfo {year} {2022})},\ \Eprint
  {http://arxiv.org/abs/1811.04968} {arXiv:1811.04968} \BibitemShut {NoStop}%
\bibitem [{\citenamefont {Katzgraber}(2011)}]{Katzgraber2009}%
  \BibitemOpen
  \bibfield  {author} {\bibinfo {author} {\bibfnamefont {H.~G.}\ \bibnamefont
  {Katzgraber}},\ }\href@noop {} {\  (\bibinfo {year} {2011})},\ \Eprint
  {http://arxiv.org/abs/0905.1629} {arXiv:0905.1629} \BibitemShut {NoStop}%
\end{thebibliography}

\appendix
\section{Low-depth QAOA simulations}
\label{app:qaoa}

Here, we provide details of our implementation of the numerical simulations of QAOA at $p=1$. If we limit ourselves to problems with general Ising-like (i.e., quadratic) cost Hamiltonians of the form 
\begin{equation}
\label{eq:sat_1-hamiltonian}
    \h{H}_c
    =
    \sum_{i} h_i \h{Z}_i
    + 
    \sum_{i<j} J_{ij}\h{Z}_i \h{Z}_j,
\end{equation}
it was shown that the expectation values of $\h{Z}_i$ and $\h{Z}_i \h{Z}_j$ according to the depth $p = 1$ QAOA state $\ket{\psi(\beta_1, \gamma_1)} =e^{- i \beta_1 \hmix} e^{- i \gamma_1 \hcost} \ket{+}^{\otimes n}$ can be calculated in $\mathcal{O}(n)$ time via simple analytical expressions~\cite{bravyi_obstacles_2020, Ozaeta2022-journal}. Because the cost Hamiltonian comprises $\mathcal{O}(n^2)$ terms, the complexity of simulating the calculation of the correlation matrix $M$ in the update step of Algorithm~\ref{alg:update-step} is $\mathcal{O}(n^3)$. Therefore, assuming $\mathcal{O}(n)$ update steps, this yields an overall complexity of $\mathcal{O}(n^4)$ for the full QIRO simulation and $\mathcal{O}(n^5)$ for the QIRO + BT simulation, when $\mathcal{O}(n)$ backtracking steps are performed, as is the case following the prescription of Algorithm~\ref{alg:qiro+bt}.

We comment here on the difference in the measurement processes between QIRO and vanilla QAOA. If the cost Hamiltonian is $k$-local, the process of finding the optimal parameters according to the expectation value of the cost Hamiltonian in principle requires only measurements with $k$ qubits involved for both algorithms. However, after the optimal parameters are found, QAOA samples classical bit strings from the quantum state for which measurements of all qubits involved are required. This makes QAOA exponentially difficult to simulate classically for any depth $p$~\cite{Farhi2016}. In contrast, the sampling step is replaced by a classical update step in QIRO. Hence, using depth $p = 1$ QAOA as state preparation for QIRO, allows us to simulate the algorithm efficiently. 

One may wonder why this should be a quantum algorithm of interest, if it is amenable to efficient classical simulations. The answer to this lies in the fact that the situation changes dramatically at higher depths of QAOA with $p > 1$ where no method for analytically computing the one-point and two-point correlations classically in polynomial time is known (to us). The same holds true at $p = 1$ when calculating the expectation values of $k$-body interaction terms, with $k\geq 3$. Therefore, these cases could lead to an algorithm which provides quantum advantage. In general, we expect an increasing performance of QIRO for higher depths of QAOA, as indicated by the MIS experiments in Section~\ref{sec:mis-results}. We note that similar observations were previously made by Bravyi~\emph{et al.}~\cite{bravyi_obstacles_2020}.

\section{Recursive QAOA (RQAOA)}
\label{app:rqaoa}

To overcome the limitations of the QAOA due to its local structure~\cite{Hastings2019, bravyi_obstacles_2020}, Bravyi~\emph{et al.}~\cite{bravyi_obstacles_2020, bravyi_hybrid_2022} introduced recursive QAOA (RQAOA). The algorithm optimizes a QAOA state based on the current problem Hamiltonian and uses this information to reduce the problem via the elimination of a single variable. Recursive application of this step reduces the problem until a specified threshold of remaining variables $n_c$ is reached. The remaining problem can be solved by a classical solver (e.g., by a brute-force search). The elimination step is motivated by the rounding of fractional correlations in relaxations of linear programs~\cite{bravyi_obstacles_2020}. While hardly a fair comparison in terms of resources required, RQAOA has been shown to significantly outperform QAOA~\cite{bravyi_hybrid_2022, bae_recursive_2022}.

It was first applied to the $\mathbb{Z}_2$-symmetric Max-Cut problem, which comprises \emph{exclusively} $2$-body terms. Here, we (trivially) extend the approach to also include single-body terms. We additionally note that a generalization to Hamiltonians with an arbitrary degree of interaction can be envisaged. However, for Hamiltonians with $K$-body terms with $K \geq 3$ one faces an additional obstacle, as the degree of the interactions can increase when an elimination step is performed. This comes with additional requirements on the quantum hardware; as we are mostly interested in NISQ-era algorithms, we do not consider such extensions here.

Assuming a quadratic Hamiltonian of the form Eq.~\eqref{eq:sat_1-hamiltonian} the pseudocode for the RQAOA, as proposed in Refs.~\cite{bravyi_obstacles_2020, bravyi_hybrid_2022}, is given in Algorithm.~\ref{alg:rqaoa}. 
\begin{algorithm}[H]
\caption{Recursive QAOA (RQAOA)}
\label{alg:rqaoa}
\hspace*{\algorithmicindent} \textbf{Input:} Problem Hamiltonian $\hat{H}$, threshold $n_\text{c}$. \\
\hspace*{\algorithmicindent} \textbf{Output} Solution $S$.
\begin{algorithmic}[1]
\State $L \gets \{\}$ \Comment{Initialize list to store frozen correlations.}
\For{$k = 1$ \textbf{to} $n - n_\text{c}$}
    \State Find optimal variational parameters of QAOA: \newline
    \hspace*{\algorithmicindent}
    $\boldsymbol{\beta}^*, \boldsymbol{\gamma}^*
    =
    \arg\,\min_{\boldsymbol{\beta}, \boldsymbol{\gamma}}
    \bra{\psi(\boldsymbol{\beta}, \boldsymbol{\gamma})}
    \hat{H}
    \ket{\psi(\boldsymbol{\beta}, \boldsymbol{\gamma})}$
    \State $\ket{\psi} \gets \ket{\psi(\boldsymbol{\beta}^*, \boldsymbol{\gamma}^*)}$
    \State Store correlations in $M\in\mathbb{R}^{n\times n}$, initialize $M_{ij}=0$: \newline 
    \hspace*{\algorithmicindent}  $\forall i \in [n]: M_{ii} = \bra{\psi} \h{Z_i}\ket{\psi}$,\newline
    \hspace*{\algorithmicindent}  $\forall (i, j) \in [n] \times [n] \text{~s.t.~} J_{ij}\neq 0: M_{ij} = \bra{\psi} \h{Z}_i \h{Z}_j \ket{\psi}$.
    \State $(i, j) = \arg\max_{(i, j) \in [n] \times [n]}|M_{ij}|$ 
    \State Append $\{(i, j), \sign(M_{ij})\}$ to $L$.
    \If{$i = j$}
    \Comment{One-point correlation.}
        \State $\h{H} \gets \text{Repl. } \h{Z}_i \text{ in } \h{H} \text{ with } \sign(M_{ii}) \cdot  \mathds{1}$
    \Else
    \Comment{Two-point correlation.}
        \State $\h{H} \gets \text{Repl. } \h{Z}_i \text{ in } \h{H} \text{ with } \sign(M_{ij}) \cdot \h{Z}_j$
    \EndIf
\EndFor

\State $S_{n_\text{c}} \gets \mathrm{Bruteforce} (\h{H})$ \Comment{Solve the remaining problem.}
\State $S \gets \mathrm{Reconstruct}(S_{n_\text{c}}, L)$ \Comment{Extract the solution from $L$.}
\Return $S$

\end{algorithmic}
\end{algorithm}

\section{MAX-2-SAT Simplification details}
\label{app:m2s-update}

Here we provide the details on the update rules employed in the QIRO update step for MAX-2-SAT. Inference rules are a commonly used building block of many classical satisfiability solvers~\cite{Li2021a}. They are used to simplify the formula by fixing the assignments of specific variables and therefore accelerate the remaining computation. Here, we use only very simple inference rules, that have low computational (and conceptual) complexity. The selected rules ensure that the optimal solutions of both the original and the simplified problem are equally good in quality. Despite their simplicity, the inference rules improve the performance of QIRO, both in terms of quality (see Section~\ref{sec:max2sat-results}), as well as in terms of resource efficiency. This is because fewer calls to the quantum device are required, due to the simplifications performed by the inference rules. The selected rules, listed next, were proposed in Refs.~\cite{Li2021a,Abrame2015}.
\begin{itemize}
    \item
    The \emph{pure literal rule}~\cite{Borchers1998}: If a literal exclusively occurs with positive (negative) polarity, then the value of the corresponding variable is set to TRUE (FALSE).
    \item
    The \emph{almost common clause rule}~\cite{Borchers1998}: If a MAX-2-SAT instance includes the clauses $x_i \lor x_j$ and $\bar{x}_i \lor x_j$, then both clauses can be replaced with a single-literal clause (i.e., unit clause) $x_j$.
    If neither $x_i$ nor $\bar{x}_i$ appears in the formula anymore, the variable can be set arbitrarily either to TRUE or FALSE.
    \item 
    The \emph{complementary unit clause rule}~\cite{Niedermeier2000}: If the formula contains single-literal clauses $x_i$ \emph{and} $\bar{x}_i$ then these two clauses are removed.
    Again, if neither $x_i$ nor $\bar{x}_i$ appears in the formula thereafter, the assignment of the variable can be set arbitrarily either to TRUE or FALSE.
    \item 
    The \emph{dominating unit clause rule}~\cite{Niedermeier2000}: If the total count of clauses, regardless of their length, that contain a variable $x_i$ ($\bar{x}_i$) is not greater than the count of unit clauses that contain $\bar{x}_i$ ($x_i$), then the variable $x_i$ is set to FALSE (TRUE).
\end{itemize}

The complete QIRO update step \texttt{Reduce}, as introduced in Section~\ref{sec:m2s-update-rules} is as follows: 
\begin{algorithm}[H]
\caption{Reduce (MAX-2-SAT)}
\label{alg:update-m2s}
\hspace*{\algorithmicindent} \textbf{Input:} Formula $\phi$, partial solution $S$. \\
\hspace*{\algorithmicindent} \textbf{Output} Simpl. formula $\phi$, updated partial solution $S$. %
\begin{algorithmic}[1]
\State Prepare low-energy quantum state $\ket{\psi}$.
\State Store correlations in $M\in\mathbb{R}^{n\times n}$:\newline%
$\forall i \in [n]: M_{ii} = \bra{\psi} \h{Z_i}\ket{\psi}$,\newline
$\forall (i, j) \in [n] \times [n] \text{~s.t.~} J_{ij}\neq 0: M_{ij} = \bra{\psi} \h{Z}_i \h{Z}_j \ket{\psi}$.
\If{$i = j$}
    \Comment{One-point correlation.}
        \State Assign variable $x_i\rightarrow \frac{1}{2}(\sign(M_{ii}) + 1)$.
\Else
\Comment{Two-point correlation.}
\If{$\sign(M_{ij})=1$}
\State Replace $x_i\rightarrow x_j$ in $\phi$ and extend $S$ accordingly.
\Else
\State Replace $x_i\rightarrow \Bar{x}_j$ in $\phi$ and extend $S$ accordingly.
\EndIf
\EndIf
\State $\phi' \gets \{\}$ \Comment{Initialize empty formula for comparison.}
\While{$\phi \neq \phi'$ and $\mathrm{size}(\phi) > n_{\mathrm{c}}$} \Comment{Inference rules.}
    \State $\phi' \gets \phi$
    \State $\phi, S \gets \text{PureLiteral}(\phi, S)$
    \State $\phi, S \gets \text{DominatingUnitClause}(\phi, S)$
    \State $\phi, S \gets \text{AlmostCommonClause}(\phi, S)$
    \State $\phi, S \gets \text{ComplementaryUnitClause}(\phi, S)$
\EndWhile
\If{$\mathrm{size}(\phi) \leq n_{\mathrm{c}}$}
\State $\phi, S \gets \text{BruteForce}(\phi, S)$
\EndIf \\
\Return $\phi$, $S$
\end{algorithmic}
\end{algorithm}

Importantly, the MAX-2-SAT QIRO update rules presented here preserve the problem structure in that only clauses with 2 or less variables appear in the formula. Consequently, the Hamiltonian remains in the same form as in Eq.~\eqref{eq:sat_1-hamiltonian}.

\section{MIS on a neutral atom quantum device}
\label{app:quera-details}

Here, we provide details about the adiabatic protocols on a neutral atom quantum processor that we used to prepare the quantum states used to guide the QIRO algorithm for the MIS problem. The MIS problem on unit disk graphs (UDGs) naturally emerges in the context of neutral atom quantum processors based on Rydberg atom arrays~\cite{ebadi_quantum_2022, pichler_quantum_2018}. UDGs are a family of graphs, where nodes are connected only if the (Euclidian) distance between is smaller than some threshold radius $R_{\mathrm{d}}$. 

Here, we generate such graphs by positioning nodes on an underlying square lattice with lattice constant $a$. We set $\sqrt{2}a<R_{\mathrm{d}}<2a$, such that we get nearest neigbhor, and diagonal connectivity. We chose a lattice constant $a=\SI{5.3}{\micro\metre}$, inspired by a previous study~\cite{finzgar_bo_2023}. The graphs presented here were generating by positioning $137$ nodes on such an underlying lattice with $14\times 14$ sites. This corresponds to a filling factor of roughly $70\%$. An example graph is shown in Figure~\ref{fig:example-graph}. 

Furthermore, we classified the generated instances in terms of their classical hardness parameter 
\begin{equation}
\label{eq:hardness-parameter}
    \mathcal{HP}:=\frac{N_{|\mathrm{MIS}|-1}}{|\mathrm{MIS}| \cdot N_{|\mathrm{MIS}|}},
\end{equation}
where $N_{M}$ denotes the number of independent sets of size $M$. The hardness parameter was shown to be related to the performance of quantum and classical algorithms alike~\cite{ebadi_quantum_2022}. However, we note that evidence has recently been put forward that $\mathcal{HP}$ only influences the performance of Markov chain Monte Carlo based algorithms (e.g., SA)~\cite{andrist_hardness_2023_1}.

Here, we consider graphs spanning several orders of magnitude in $\mathcal{HP}$, which we computed using the tropical tensor network algorithm provided within the \texttt{GenericTensorNetworks} library~\cite{liu_computing_2023, liu_tropical_2021}. Specifically, the graphs considered in Fig.~\ref{fig:quera-data} have been sorted in order of increasing hardness, with the exact values of $\mathcal{HP}$ given in Table~\ref{tab:hp}. 

\begin{table}[h]
\caption{Values of the hardness parameter $\mathcal{HP}$ for the graph instances presented in Fig.~\ref{fig:quera-data}.}
\begin{tabular*}{\columnwidth}{@{\extracolsep{\fill}} l c c c c c c c c c c}
\hline \hline
Graph instance & 1 & 2 & 3 & 4 & 5 & 6 & 7 & 8 & 9 & 10 \\ \hline
$\mathcal{HP}$ & 3 & 8 & 14 & 39 & 49 & 95 & 197 & 383 & 863 & 1435 \\
\hline \hline
\end{tabular*}
\label{tab:hp}
\end{table}

We now briefly summarize how neutral atom arrays can be used to solve the MIS problem on UDGs~\cite{pichler_quantum_2018, ebadi_quantum_2022}. The dynamics of neutral atom quantum devices are governed by the Hamiltonian $\h{H}(t)=\h{H}_{\mathrm{dr}} + \h{H}_{\mathrm{cost}}$, where the individual terms read:
\begin{equation}
\label{eq:rydberg-hamiltonian}
\begin{aligned}
    \hhat_{\mathrm{dr}} \,\,\,\,
    &=
    \frac{\hbar}{2} \Omega(t) \sum_i
    \h{X}_i
    \\
    \hhat_{\mathrm{cost}}
    &=
    -\hbar \Delta(t) \sum_i \hat{n}_i
    +
    \sum_{i<j} V_{i j} \hat{n}_i \hat{n}_j.
\end{aligned}
\end{equation}
Here, $\Omega$ is the Rabi frequency, $\Delta$ is the laser detuning, and the interaction term $V_{ij}=C_6\lVert \mathbf{x}_i - \mathbf{x}_j \lVert_2^{-6}$, with $C_6$ the van der Waals coefficient. The particular value $C_6$ depends on the atom species used in the experimental setup. The number operator $\hat{n}_i:=\frac{1}{2}(\mathds{1}_2-\h{Z}_i)$ counts the number of Rydberg excitations on the $i$-th site.

\begin{figure}[htb]
\includegraphics{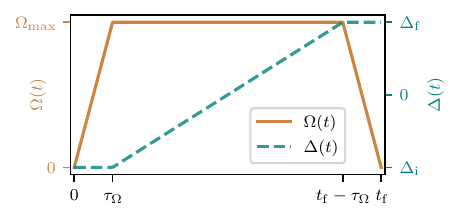}
\caption{Time dependency of the Rabi frequency $\Omega(t)$ and the laser detuning $\Delta(t)$. We use $t_{\mathrm{f}}=\SI{4}{\micro\second}$, $\Omega_{\max}=\SI{15.8}{\mega\hertz}$, $\tau_{\Omega}=0.1 t_{\mathrm{f}}$, $\Delta_{\mathrm{i}}=\SI{-30}{\mega\hertz}$ and $\Delta_{\mathrm{f}}=\SI{60}{\mega\hertz}$.}
\label{fig:rydberg-schedules}
\end{figure}

An essential property of the Hamiltonian in Eq.~\eqref{eq:rydberg-hamiltonian} is the so-called Rydberg blockade phenomenon, in which two atoms cannot simultaneously be in the (excited) state $\ket{1}$ if the distance between them is smaller than the Rydberg blockade radius $R_{\mathrm{b}} \equiv \left(
C_6/
\hbar\Omega
\right)^{1/6}$~\cite{ebadi_quantum_2022}.
Therefore, if one sets the UDG radius to be equal to the Rydberg blockade radius ($R_{\mathrm{d}}=R_{\mathrm{b}}$), ground states of $\hhat_\mathrm{cost}$ obey the independence constraint on UDGs if one assigns atoms that are in the $\ket{1}$ to the MIS~\cite{pichler_quantum_2018}. Additionally, if we set $0 < \Delta < V_{ij}$, the ground state of $H_\mathrm{cost}$ maximizes the number of excitations, while not violating independence, hence corresponding to solutions to the original MIS problem. We thus start the adiabatic protocol with $\Delta_{\mathrm{i}} < 0$, such that the ground state is $\ket{0}^{\otimes |V|}$ and then (slowly) increase the detuning until some final positive value, as shown in Fig.~\ref{fig:rydberg-schedules}, such that the ground states of the final Hamiltonian correspond to the solutions of the MIS problem.

We ran such protocols on the QuEra Aquila device, which is available through the Amazon Braket service. The detailed parameters can be found in the caption of Fig.~\ref{fig:rydberg-schedules}, and are partially motivated by a previous study~\cite{finzgar_bo_2023}. At each stage of the QIRO algorithm we estimate the correlations using a hundred measurements of the quantum state prepared by the adiabatic protocol described in Figure~\ref{fig:rydberg-schedules}.

\section{Higher depth QAOA experiments}
\label{app:higher-depth-qaoa}
To provide additional support to the claim that a higher quality of correlations yields better performance of QIRO, we performed numerical simulations of QIRO using correlations from QAOA at $p>1$. In contrast to $p=1$, no analytical formulae for the required correlators are known at $p>1$, forcing us to resort to state-vector simulations of QAOA circuits. Consequently, we had to limit our experiments to solving the MIS problem on smaller graph instances with $12$ nodes, randomly chosen from the Erd\H{o}s-R\'enyi ensemble as in Section~\ref{sec:mis-results}. For each fixed expected degree, we generated a hundred random instances, and found their MIS using a brute force search of the solution space. We then ran QIRO informed by correlations from the QAOA implementation in the \texttt{PennyLane} library~\cite{bergholm2022pennylane}, using the gradient descent optimizer with 300 iterations and 15 restarts of the optimizer at different random initial values of the variational parameters. For these experiments we set $n_{\mathrm{c}}=1$ because of the comparatively smaller problem instance sizes.

\begin{figure}[htb]
\includegraphics{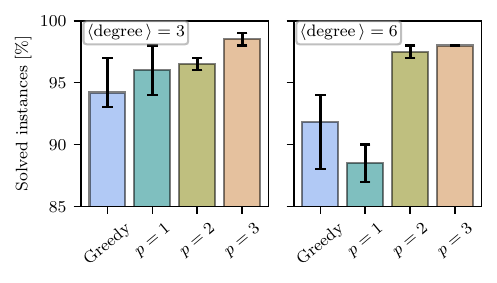}
\caption{The fraction of optimally solved instances by QIRO using QAOA at depths $p\in\{1,2,3\}$, and by the minimal degree greedy algorithm. The height of the bars show the mean fraction across ten runs of the algorithms, with the error bars indicating the best and worst performance across those runs.} 
\label{fig:higher-p}
\end{figure}

The results in Fig.~\ref{fig:higher-p} show the fraction of optimal independent sets found by QIRO with correlations from QAOA at depths $p\in\{1,2,3\}$, and by the minimal degree greedy algorithm. The performance of QIRO clearly improves with increased QAOA depth, for both considered graph densities. Together with the results in Section~\ref{sec:mis-results}, this provides yet another indication that enhanced correlations lead to better performance of QIRO. Moreover, these experiments indicate that correlations from deeper circuits enable QIRO to surpass the performance of the greedy benchmark. Although this matches the intuitive expectations, further experiments on larger instances and quantum hardware are required to strengthen these claims.

\section{Further RQAOA validity experiments}
\label{app:rqaoa-validity}

In Section~\ref{sec:mis-results} we assessed the performance of QIRO and RQAOA at different qualities of QAOA parameters. As the results presented in Fig.~\ref{fig:mis-validity-apprat} exhibited interesting behavior, we here report an extended set of experiments, where RQAOA validities for additional magnitudes of the penalty term $\lambda$ [cf. Eq.~\eqref{eq:mis-hamiltonian}] are shown.

Interestingly, the results partially defy the na\"{\i}ve expectation that a larger penalty term should favor enforcing constraints. This is especially prominent in the region between $\lambda=1.2$ and $\lambda=5$, where the robustness of RQAOA seems to decrease. However, the trend reverses again for $\lambda=10$.

\begin{figure}[htb]
\includegraphics{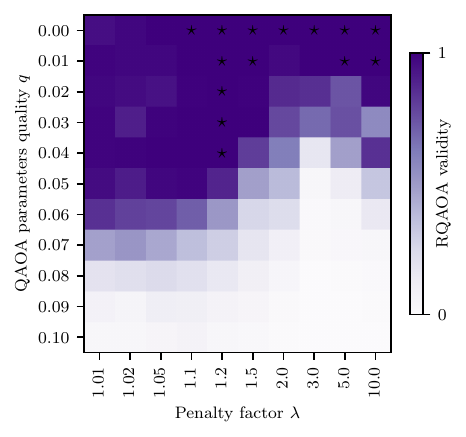}
\caption{Ratio of valid solutions obtained by RQAOA for different qualities of the QAOA parameters $q$ (at $p=1$) and different penalty factors $\lambda$. As in the results in Fig.~\ref{fig:mis-validity-apprat}, at each $(q, \lambda)$ the same fifty UDGs of size 137 were generated and ten runs of RQAOA were performed. We here report the ratio valid (i.e., independent) sets found by RQAOA. Black stars indicate values of $(q, \lambda)$ where all found solutions were feasible.} 
\label{fig:validity-matrix}
\end{figure}

While a thorough analysis of the artefacts observed in Fig.~\ref{fig:validity-matrix} is beyond the scope of this paper, we offer some insights into what might underpin the peculiar behavior at intermediate values of $\lambda$. We suspect that this is caused by an intricate interplay between the deformations of the spectrum of the cost Hamiltonian Eq.~\eqref{eq:mis-hamiltonian} when $\lambda$ is varied, with the particularities of QAOA states at $p=1$. As $\lambda$ is varied, only the energy levels of states that correspond to configurations with violations of the independence constraint are changed. Thus, the nature of the excited states can change dramatically, i.e., states corresponding to configurations with violations might shift relative to states corresponding to independent sets. It is difficult to predict the amplitudes of such excited states in the optimal QAOA states at $p=1$, let alone at perturbed parameters. As such, a pattern as the one seen is not inconceivable. However, more experiments are required to pin down the exact underlying mechanism.

\section{Parallel tempering and simulated annealing setup}
\label{app:pt-sa-setup}

For a MAX-2-SAT instance with $n$ variables, starting from a random initial configuration, we perform simulated annealing with $600 \cdot n$ attempted variable flips. The annealing schedule consists of evenly spaced inverse temperatures $\beta = T^{-1}$, with an initial value $\beta = 0$ and a final inverse temperature of $\beta = 6$.

For parallel tempering (PT)~\cite{Hukushima_1996,Earl2005, Roma2009} we initialize $12$ replicas of the problem in random configurations with different temperatures. We chose the following fine-tuned the temperatures: 
\begin{table}[H]
\caption{Temperatures of the replicas in our implementation of PT.}
\begin{tabular*}{\columnwidth}{@{\extracolsep{\fill}} c c c c c c c c c c c c}
\hline \hline
1 & 2 & 3 & 4 & 5 & 6 & 7 & 8 & 9 & 10 & 11 & 12 \\ \hline
0.10 & 0.20 & 0.29 & 0.39 & 0.50 & 0.62 & 0.75 & 0.90 & 1.09 & 1.33 & 1.67 & 2.20\\
\hline \hline
\end{tabular*}
\label{tab:pt-temps}
\end{table}
\noindent These were chosen in accordance with the prescription that replica exchanges should be accepted with probabilities between $0.2$ and $0.8$~\cite{Katzgraber2009}. We tested this empirically for a collection of representative MAX-2-SAT instances. We then used the following PT scheme. In each run of PT we carried out $15000$ cycles, with one simulation cycle consisting of one Monte Carlo sweep ($n$ single spin flips) per replica followed by an attempted replica exchange between all replica pairs with neighboring temperatures, starting from the lowest temperature. The configuration with the lowest energy found in any of the replicas during the course of the algorithm was finally returned as the solution.

\end{document}